%% file: main.tex
\documentclass[amsmath,amssymb,aps,pra,reprint,superscriptaddress,longbibliography]{revtex4-2}

\usepackage[svgnames]{xcolor}
\usepackage{tabularx}
\usepackage{floatrow}
\usepackage{graphicx}
\usepackage[label font=large]{subfig}
\usepackage{url}
\usepackage{braket}
\usepackage{amsmath,amssymb}
\usepackage{graphicx}
\graphicspath{{.}{figs/}{../figs/}}
\usepackage{afterpage}
\usepackage{dcolumn}
\usepackage{bm}
\usepackage{tikz}
\usepackage{quantikz}
\usetikzlibrary{positioning,intersections,angles,quotes}
\usepackage{comment}
\usepackage{physics}
\usepackage{mathtools}
\usepackage{chemfig}
\usepackage[version=4]{mhchem}
\usepackage[colorlinks,citecolor=DarkGreen,linkcolor=FireBrick,linktocpage,unicode,hypertexnames=false]{hyperref}

\usepackage{ulem}
\usepackage{adjustbox}

\makeatletter
\def\@opargbegintheorem#1#2#3{\trivlist
   \item[]{\bfseries #1\ #2\ (#3)} \itshape}
\makeatother

\DeclareFloatSeparators{mysep}{\hskip-30em}

\usepackage{enumerate}

\begin{document}

\title{Symmetry-Adapted State Preparation for Quantum Chemistry on Fault-Tolerant Quantum Computers}

\author{Viktor Khinevich}%
\email{victorkh711@gmail.com}
\affiliation{%
  Graduate School of Engineering Science, The University of Osaka, 1-3 Machikaneyama, Toyonaka, Osaka 560-8531, Japan
}%
\affiliation{%
  Center for Quantum Information and Quantum Biology,
  The University of Osaka, 1-2 Machikaneyama, Toyonaka 560-8531, Japan
}%

\author{Wataru Mizukami}%
\email{mizukami.wataru.qiqb@osaka-u.ac.jp}
\affiliation{%
  Graduate School of Engineering Science, The University of Osaka, 1-3 Machikaneyama, Toyonaka, Osaka 560-8531, Japan
}%
\affiliation{%
  Center for Quantum Information and Quantum Biology,
  The University of Osaka, 1-2 Machikaneyama, Toyonaka 560-8531, Japan
}%

	
\begin{abstract}
We present systematic and resource-efficient constructions of continuous symmetry projectors, particularly $U(1)$ particle number and $SU(2)$ total spin, tailored for fault-tolerant quantum computations.
Our approach employs a linear combination of unitaries (LCU) as well as generalized quantum signal processing (GQSP and GQSVT) to implement projectors. These projectors can then be coherently applied as state filters prior to quantum phase estimation (QPE). 
We analyze their asymptotic gate complexities for explicit circuit realizations.
For the particle number and $S_z$ symmetries, GQSP offers favorable resource usage features owing to its low ancilla qubit requirements and robustness to finite precision rotation gate synthesis.
For the total spin projection, the structured decomposition of $\hat{P}_{S,M_S}$ reduces the projector T gate count.
Numerical simulations show that symmetry filtering substantially increases the QPE success probability, leading to a lower overall cost compared to that of unfiltered approaches across representative molecular systems.
Resource estimates further indicate that the cost of symmetry filtering is $3$ to $4$ orders of magnitude lower than that of the subsequent phase estimation step
This advantage is especially relevant in large, strongly correlated systems, such as FeMoco, a standard strongly correlated open-shell benchmark.
For FeMoco, the QPE cost is estimated at ${\sim}10^{10}$ T gates, while our symmetry projector requires only ${\sim}10^{6}$--$10^{7}$ T gates.
These results establish continuous-symmetry projectors as practical and scalable tools for state preparation in quantum chemistry and provide a pathway toward realizing more efficient fault-tolerant quantum simulations.
\end{abstract}

\maketitle

\input{Sections/01_Introduction}

\input{Sections/02_Theory}

\input{Sections/03_Computational_details}

\input{Sections/04_Results}
\input{Sections/05_Conclusions}

\begin{acknowledgments}
This study was supported by the MEXT Quantum Leap Flagship Program (MEXTQLEAP) Grant No. JPMXS0120319794; JST COI-NEXT Program, Grant No. JPMJPF2014 and JST ASPIRE Program JPMJAP2319.
This research was partially supported by the JSPS Grants-in-Aid for Scientific Research (KAKENHI) Grant No. JP23H03819.
We thank the Supercomputer Center, the Institute for Solid State Physics, the University of Tokyo for the use of the facilities. 
This work was also conducted using SQUID at the Cybermedia Center, the University of Osaka.
\end{acknowledgments}

\input{Sections/06_Appendix}

\bibliography{main.bib}

\end{document}

%% file: Sections/01_Introduction.tex
\section{Introduction} 
\label{sec:introduction}

Quantum computing has advanced significantly in recent years, supported by progress in hardware platforms and the design of quantum algorithms that can address classically intractable problems~\cite{AbuGhanem2025,Jordan2025,Acharya2025,Moses2023}. 
Among the most promising applications is the simulation of quantum many-body systems, particularly in quantum chemistry, where the ability to describe electronic structure accurately is directly related to the understanding of chemical reaction mechanisms, material properties, and molecular spectroscopy~\cite{McArdle2020RMP,Bauer2020ChemRev,Huh2015Vibronic,Colless2018PRX,kunitsa2025quantumsimulationelectronenergy}. 
Although many quantum algorithms have been proposed to compute ground- and excited-state energies~\cite{AbramsLloyd1999QPE,Peruzzo2014VQE,McClean2017QSE,Nakanishi2019SSVQE,kanno2023quantumselectedconfigurationinteractionclassical}, a recurring challenge is the preparation of high-fidelity initial states suitable for such computations.

State preparation plays a central role in algorithms such as quantum phase estimation (QPE)~\cite{kitaev1995quantummeasurementsabelianstabilizer,AbramsLloyd1999QPE}, where the probability of successful projection onto an eigenstate is determined by the overlap between the prepared input state and the desired eigenstate of the target Hamiltonian. 
In practice, the overlap is often limited when generic or heuristic initial states are employed. 
Various methods have been explored to improve the state preparation, including variational optimization~\cite{Peruzzo2014VQE, Kandala2017,Romero2018UCC, Grimsley2019ADAPT}, adiabatic state preparation~\cite{AspuruGuzik2005ASP,Veis2014ASP}, tensor networks~\cite{Schon2005,Malz2024MPS, Berry2025}, and state filters~\cite{lee2025filteredquantumphaseestimation, ding2025quantumfilteringanalysismultiplicities, sakuma2025quantumphaseestimationbased, Lin2020optimalpolynomial, Irmejs2024efficientquantum}. 
However, these strategies can suffer from limitations such as barren plateau behavior, large circuit depths, or the need for classical preoptimization.

An alternative approach uses the inherent symmetries of physical Hamiltonians. 
Molecular Hamiltonians, in particular, possess well-defined symmetry sectors and physical eigenstates reside within specific irreducible representations of the corresponding symmetry group. 
Projecting an initial state onto a chosen symmetry sector can significantly enhance the overlap with the targeted eigenstates and reduce the effective Hilbert space explored by the algorithm. 

The traditional use of symmetries may require non-scalable resources or rely on repetitive measurements, limiting their applicability in fault-tolerant settings~\cite{Lacroix2023, Tsuchimochi2020}.
One method to optimize the state of a particular symmetry sector is to use penalty terms~\cite{Ryabinkin2019}.
Penalty-based methods usually introduce many additional terms into the Hamiltonian to measure, and require parameter fitting.
QPE can also be used for symmetry-state filtering~\cite{Siwach2021}.
Another approach uses Löwdin spin projectors to modify the Hamiltonian~\cite{Yen2019}.
This leads to an exponential number of terms, making its implementation impractical.
However, in this study, we demonstrate how to implement this type of projector efficiently.
There is a recent general approach for constructing any symmetry projector, it is limited to finite-symmetry groups, such as point groups~\cite{Bastidas2025}.

In this study, we constructed projectors for continuous symmetries relevant to the electronic structure, including the particle number and spin projection ($U(1)$) and total spin ($SU(2)$), and demonstrate how to implement Löwdin-type projectors efficiently.
We implemented them using a linear combination of unitaries (LCU)~\cite{LCU} and generalized quantum signal processing~\cite{GQSP} (GQSP and GQSVT~\cite{GQSVT}) frameworks.
We analyzed the theoretical scaling of these constructions under fault-tolerant quantum computing (FTQC) assumptions. We also evaluated their numerical accuracy and sensitivity to the finite precision of rotational gate synthesis.
We further provide detailed resource estimates demonstrating their practicality as state filters for QPE in molecular Hamiltonian simulations, showing that symmetry projectors enhance state overlap in realistic quantum chemistry settings.
Finally, we estimate the required cost for the FeMoco complex, a standard benchmark for strongly correlated open-shell systems~\cite{Reiher2017,Li2019}. 

The remainder of this paper is organized as follows. 
Sec.~\ref{sec:theory} presents the theoretical background and formal construction of the symmetry projectors, including their properties.
Sec.~\ref{sec:computation} describes the computational stack and provides source code for reproducibility. 
Sec.~\ref{sec:results} reports theoretical scaling and identifies the optimal parameter regimes, followed by numerical calculations and resource estimates with representative molecular case studies. 
This section outlines the open challenges and prospective improvements of this methodology. 
Finally, Sec.~\ref{sec:conclusion} summarizes the main contributions and discusses future directions.

%% file: Sections/02_Theory.tex
\section{Theory} 
\label{sec:theory}

\subsection{Symmetry operators} 
\label{subsec:symmetry_operators}

We aimed to prepare the quantum states in the fixed-symmetry sector of the electronic Hamiltonian.
We denote $\hat{N}$, $\hat{S}_z$, and $\hat{S}^2$ as the particle number, spin projection onto the selected $z$-axis, and total spin-squared operators, respectively.
Each operator is Hermitian and we aim to prepare quantum states belonging to the common subspace defined by 
\begin{equation}
\begin{aligned}
\hat{N} |\psi \rangle &= N_\text{elec} |\psi\rangle,\\
\hat{S}_z |\psi \rangle &= M_S |\psi\rangle,\\
\hat{S}^2 |\psi \rangle &= S(S+1) |\psi\rangle,
\end{aligned}
\end{equation}
where $N_\text{elec}$ is the number of particles (electrons in this case), $M_S$ is the spin projection, and $S$ is the total spin quantum number.

To express these operators in the second quantized form, we adhere to the convention that even spin-orbital indices correspond to the spin state $|\alpha\rangle$ (or $|\uparrow\rangle$), and odd indices correspond to spin $|\beta\rangle$ (or $|\downarrow\rangle$).
Indexing starts at $0$.
The particle number operator is expressed as
\begin{equation}
\label{eq:N_op_sq}
    \hat{N} = \sum_{i=0}^{N_\text{SO}-1}  \hat{a}_i^\dagger  \hat{a}_i, 
\end{equation}
where $\hat{a}_i^\dagger$ and $\hat{a}_i$ are fermionic creation and annihilation operators acting on spin-orbital $i$, $N_\text{SO}$ is the number of spin-orbitals.

Similarly, the spin-projection operator is
\begin{equation}
\label{eq:Sz_op_sq}
    \hat{S}_z = \frac{1}{2}  \sum_{i=0}^{N_\text{SO}-1} (-1)^{i} \hat{a}_i^\dagger  \hat{a}_i.
\end{equation}

The second quantized form of the total spin-squared operator $\hat{S}^2$ is more complicated and requires the introduction of spin ladder operators:
\begin{equation}
\label{eq:ladder_ops}
    \hat{S}_+ = \sum_{i=0}^{N_\text{SO}/2-1} \hat{a}_{2i}^\dagger  \hat{a}_{2i+1}, \quad 
    \hat{S}_- = \sum_{i=0}^{N_\text{SO}/2-1} \hat{a}_{2i+1}^\dagger  \hat{a}_{2i}.
\end{equation}
The total spin operator can then be written as:
\begin{equation}
\label{eq:S2_op_sq}
    \hat{S}^2 = \hat{S}_z^2 + \hat{S}_z + \hat{S}_- \hat{S}_+
              = \hat{S}_z^2 - \hat{S}_z + \hat{S}_+ \hat{S}_-.
\end{equation}

To use these operators on a quantum computer, we must map the fermionic operators to qubit operators via a fermion-to-qubit transformation.
Let the fermionic system be encoded onto $N_\text{SO}$ qubits using the Jordan–Wigner transformation.
In this mapping, the number operator becomes
\begin{equation}
\label{eq:N_op_JW}
    \hat{N} \xrightarrow{\text{JW}} 
    \sum_{j=0}^{N_\text{SO}-1} \frac{I - Z_j}{2} 
    = \frac{N_\text{SO}}{2}I - \frac{1}{2}\sum_{j=0}^{N_\text{SO}-1} Z_j.
\end{equation}
Thus, the number operator is a linear combination of Pauli strings with a total number of terms $L = N_\text{SO} + 1$.

Similarly, the following simple expression holds for $\hat{S}_z$:
\begin{equation}
\label{eq:Sz_op_JW}
    \hat{S}_z \xrightarrow{\text{JW}} 
    \frac{1}{4} \sum_{p=0}^{N_\text{SO}/2-1} (Z_{2p+1} - Z_{2p}).
\end{equation}
For $\hat{S}_z$, the number of distinct Pauli terms was $L = N_\text{SO}$.

The Jordan–Wigner image of $\hat{S}^2$ is more complex.  
By introducing the local spin-projection operator $D_p := Z_{2p+1} - Z_{2p}$ and ladder operators $\sigma^{\pm} = (X \pm iY)/2$, we obtain the following:
\begin{equation}
\label{eq:S2_op_JW}
\begin{aligned}
    \hat{S}^2 &\xrightarrow{\text{JW}} \;
    \sum_{p=0}^{N_\text{SO}/2-1} \tfrac{3}{8}(1 - Z_{2p}Z_{2p+1})  \\[4pt]
    &+ \sum_{p<q}^{N_\text{SO}/2-1} \Bigg[
        \tfrac{1}{8} D_p D_q 
        + \big(
            \sigma_{2p}^+ \sigma_{2p+1}^- 
            \sigma_{2q}^- \sigma_{2q+1}^+ \\[-2pt]
    &\hspace{7em}
            +\;
            \sigma_{2p}^-\sigma_{2p+1}^+ 
            \sigma_{2q}^+ \sigma_{2q+1}^- 
        \big)
    \Bigg].
\end{aligned}
\end{equation}
The number of terms in this expansion scales as $L = \frac{1}{2}(3N_\text{SO}^2 - 5N_\text{SO}) + 1$, which is quadratic with $N_\text{SO}$.

Although the exact eigenstates of the electronic Hamiltonian respect these symmetries, variational ansatzes and state preparation techniques in quantum computing often break them down. 
Therefore, explicit projection onto the desired symmetry sectors can substantially improve the quality of the approximate states and the accuracy of the measurements.

\subsection{Symmetry projectors} 
\label{subsec:symmetry_projectors}

By definition, orthogonal projectors are operators satisfying $\hat{P}^2 = \hat{P} = \hat{P}^\dagger$. 
In the present context, we seek operators that project an arbitrary state $|\psi\rangle$ onto the eigenspace of a symmetry operator $\hat{O}$ corresponding to a specific eigenvalue $o_i$:
\begin{equation}
    \hat{O}\,\hat{P}_{o_i}|\psi\rangle = o_i\,\hat{P}_{o_i}|\psi\rangle.
\end{equation}
We assume that the initial state $|\psi\rangle$ exhibits a nonzero overlap with the target eigenspace.

Such a projector can be constructed using several methods~\cite{Yen2019}.
In the following, we outline the approaches used in this study, which are compatible with modern quantum computing frameworks, such as LCU, GQSP, and GQSVT (see Appendix~\ref{appendix:primitives}).

\subsubsection{Integral projectors}
\label{subsubsec:integral}
For operators $\hat{N}$ and $\hat{S}_z$, the exact projector onto eigenvalue $o_i$ can be expressed as a Fourier-type $U(1)$ group integral over a unit circle:
\begin{equation}
\label{eq:proj_int}
    \hat{P}_{o_i} = \frac{1}{2\pi} \int_0^{2\pi} e^{i\phi(o_i - \hat{O})} \, d\phi,
\end{equation}
where $\hat{O}$ is the symmetry operator, and $o_i$ is the eigenvalue corresponding to the desired subspace.

In practice, this continuous integral is replaced by a discrete Fourier sum, resulting in a finite linear combination of unitary operators as explained in Sec.~\ref{subsec:implementation}.

The projector onto a definite total-spin subspace $S$ with projection $M_S$ can be written as a $SU(2)$ group integral:
\begin{equation}
\label{eq:proj_int_S2}
\begin{aligned}
    \hat{P}_{S, M_S} = \frac{2S+1}{8\pi^2}
    &\int_0^{2\pi} d\alpha
     \int_0^{\pi} d\beta\,\sin\beta
     \int_0^{2\pi} d\gamma \; \\
    &\times D^{S}_{M_SM_S}(\alpha,\beta,\gamma)^{*}
     R(\alpha,\beta,\gamma).
\end{aligned}
\end{equation}
where $R(\alpha,\beta,\gamma) = e^{-i\alpha \hat{S}_z} e^{-i\beta \hat{S}_y} e^{-i\gamma \hat{S}_z}$ is a spin-rotation operator expressed in terms of the Euler angles.
$D^{S}_{M_SM_S}(\alpha,\beta,\gamma)$ is the corresponding Wigner $D$-matrix element.

This expression can be rewritten as follows:
\begin{equation}
\label{eq:proj_int_S2_PPP}
\begin{aligned}
    &\hat{P}_{S, M_S} = \frac{1}{2\pi}
    \int_0^{2\pi} e^{i(M_S-\hat{S}_z) \alpha} d\alpha \\
     &\times \frac{2S+1}{2} \int_0^{\pi} e^{-i\beta \hat S_y} d^{\,S}_{M_SM_S}(\beta)^*\sin\beta d\beta \\
     &\times\frac{1}{2\pi} \int_0^{2\pi} e^{i(M_S-\hat{S}_z)\gamma} d\gamma =\\
     &\hat{P}_{M_S} \Bigg[\frac{2S+1}{2} \int_0^{\pi} e^{-i\beta \hat S_y} d^{\,S}_{M_SM_S}(\beta)^*\sin\beta d\beta \Bigg] \hat{P}_{M_S},
\end{aligned}
\end{equation}
where $d^{S}_{M_SM_S}(\beta) = e^{iM_S\alpha} D^{S}_{M_SM_S}(\alpha,\beta,\gamma) e^{iM_S\gamma}$ denotes the reduced Wigner $d$-matrix. 
The integral in square brackets we will call a $\hat{P}_S$ projector.

If the initial wave function is an eigenfunction of $\hat{S}_z$ with eigenvalue $M_S$, which is the case for some ansatzes such as UCC~\cite{Romero2018UCC}, then the first $\hat{P}_{M_S}$ projector is a trivial identity.
Thus, we only need to apply a reduced $\hat{P}_S$ projector followed by a $\hat{P}_{M_S}$ projector.
This construction provides a computational advantage to spin-conserving ansatzes. 

As in Eq.~\eqref{eq:proj_int}, the integrals must be discretized for implementation. 
In the case of $\hat{P}_{S, M_S}$ and $\hat{P}_{S}$ the quadrature nodes and weights are generally nonuniform, limiting the direct applicability of GQSP.
Nevertheless, the resulting discretization yields an LCU-compatible form (see Sec.~\ref{subsec:implementation}).

\subsubsection{Lagrange interpolation projectors}

An alternative and fully algebraic approach for constructing projectors is based on the Lagrange interpolation formula.
The objective is to construct a polynomial function of the symmetry operator $\hat{O}$ that nullifies all the undesired eigenvalues $o_j$ $(j \neq i)$ while retaining the target eigenvalue $o_i$. 
The general expression is as follows:
\begin{equation}
\label{eq:lagrange_proj}
    \hat{P}_{o_i} = 
    \prod_{o_i \neq o_j} 
    \frac{\hat{O} - o_j}{o_i - o_j}.
\end{equation}
This polynomial depends solely on the Hermitian operator $\hat{O}$ and therefore can be efficiently implemented using GQSVT. 
This provides an exact construction of an arbitrary symmetry operator.
In particular, by using this approach, the Löwdin projector can be constructed for the operator $\hat{S}^2$~\cite{lowdin}, which will be used in future sections.

Note that because we work on a finite spin-orbital basis, all the symmetry operators $\hat{O}$ are finite Hermitian matrices.
Consequently, the integral projector~\eqref{eq:proj_int} and Lagrange interpolation projector~\eqref{eq:lagrange_proj} represent the same polynomial function as in the Cayley–Hamilton theorem.
This equivalence holds only when a single eigenvalue $o_i$ is targeted and the number of quadrature nodes in the integral is sufficiently large.
Despite their mathematical equivalence, these two expressions allow distinct and potentially advantageous implementations on quantum computers.

\subsection{Projectors implementation} 
\label{subsec:implementation}

\subsubsection{Linear combination of unitaries}

\noindent\textbf{$\hat{N}$ and $\hat{S}_z$ operators.}

The projectors onto the selected symmetry subspaces for $\hat{N}$ and $\hat{S}_z$ can be constructed using the integral representation introduced in Sec.~\ref{subsec:symmetry_projectors}. 
In practice, this integral is replaced by a finite Fourier sum over discrete phase points.
\begin{equation}
\label{eq:finite_sum_proj}
    \hat{P}_{o_i} = \frac{1}{N_\phi}\sum_{k=0}^{N_\phi-1} 
    e^{i \phi_k (o_i - \hat{O})}, \qquad 
    \phi_k = \frac{2\pi k}{N_\phi},
\end{equation}
where $N_\phi$ is selected to resolve all the distinct eigenvalues of $\hat{O}$.  
For the number operator $\hat{N}$ acting on $N_\text{SO}$ qubits, $N_\text{SO}+1$ distinct eigenvalues exist, and thus, $N_\phi = N_\text{SO}+1$ integration nodes are sufficient for an exact representation.

The exponential terms in Eq.~\eqref{eq:finite_sum_proj} is a unitary operator of the form $e^{-i \phi_k \hat{O}}$ with the corresponding phase $e^{i \phi_k o_i}$.  
The resulting linear combination of unitaries can be implemented using a standard LCU framework with an ancillary register comprising $\lceil\log_2 N_\phi\rceil$ qubits (see Appendix~\ref{appendix:primitives}). 
The controlled unitaries corresponding to different values $\phi_k$ are conditionally applied, based on the ancillary register state.

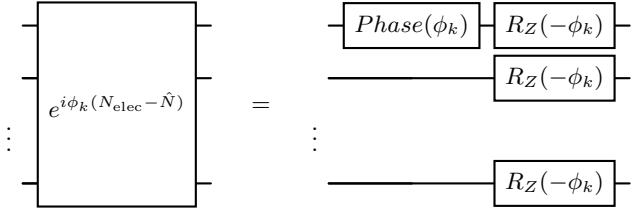
\begin{figure}[h]
    \centering
    \begin{quantikz}[column sep=0.2cm,row sep={0.7cm,between origins}]
        \lstick{} & \gate[4]{e^{i \phi_k (N_\text{elec} - \hat{N})}} &\\ 
        \lstick{} & \qw &\\
        \lstick{$\vdots$} \\
        \lstick{} & \qw &
    \end{quantikz}
    \quad $=$ \quad
    \begin{quantikz}[column sep=0.2cm,row sep={0.7cm,between origins}]
        \lstick{} & \gate{Phase(\phi_k)} & \gate{R_Z(-\phi_k)} &\\
        \lstick{} & \qw & \gate{R_Z(-\phi_k)} &\\
        \lstick{$\vdots$} \\
        \lstick{} & \qw & \gate{R_Z(-\phi_k)} & 
    \end{quantikz}
    \caption{Decomposition of the unitary operator $e^{i \phi_k (N_\text{elec} - \hat{N})}$ into single-qubit $R_Z$ rotations and a phase gate $Ie^{i\phi_k (N_\text{elec}-N_\text{SO}/2)}$ on the first qubit.}
    \label{fig:expN}
\end{figure}

The unitaries $e^{i \phi_k (N_\text{elec} - \hat{N})}$ for $\hat{P}_N$ projector can be constructed using Eq.~\eqref{eq:N_op_JW}.
These unitaries are diagonal on a computational basis and can be implemented efficiently using single-qubit $R_Z$ rotations (see Fig.~\ref{fig:expN}).

\begin{figure}[h]
    \centering
    \begin{quantikz}[column sep=0.1cm,row sep={0.7cm,between origins}]
        \lstick{} & \gate[4]{e^{i \phi_k (M_S - \hat{S}_z)}} &\\ 
        \lstick{} & \qw &\\
        \lstick{$\vdots$} \\
        \lstick{} & \qw &
    \end{quantikz}
    \quad $=$ \quad
    \begin{quantikz}[column sep=0.2cm,row sep={0.7cm,between origins}]
        \lstick{} & \gate{Phase(\phi_k)} & \gate{R_Z(-\phi_k/2)} &\\
        \lstick{} & \qw & \gate{R_Z(+\phi_k/2)} &\\
        \lstick{$\vdots$} \\
        \lstick{} & \qw & \gate{R_Z(+\phi_k/2)} &
    \end{quantikz}
    \caption{Decomposition of the unitary operator $e^{i \phi_k (M_S - \hat{S}_z)}$ into single-qubit $R_Z$ rotations and a phase gate $Ie^{i\phi_k M_S}$ on the first qubit.}
    \label{fig:expSz}
\end{figure}
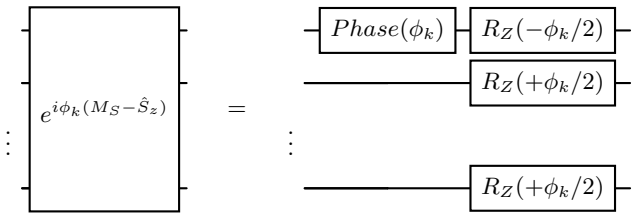

The same construction applies to $\hat{P}_{M_S}$ spin projectors, differing only in the form of unitary exponentials $e^{-i\phi_k \hat{S}_z}$.
This expression can be obtained from Eq.~\eqref{eq:Sz_op_JW}, which corresponds to the relative phase rotations between $\alpha$ and $\beta$ spin-orbitals, as shown in Fig.~\ref{fig:expSz}.

\begin{figure}[h]
    \centering
    \begin{quantikz}[column sep=0.5cm, row sep={0.5cm}]
        \lstick{$|0\rangle$} & \qwbundle{n} & \gate{\operatorname{Had}^{\otimes n}} & \gate[2]{\operatorname{SELECT}} & \gate{\operatorname{Had}^{\otimes n}} & \meter{|0\rangle}\\
        \lstick{$|\psi\rangle$} & \qw & \qw & \qw & \arrow[r] & \rstick{$\hat{P}_{o_i}|\psi\rangle$}
    \end{quantikz}
    \caption{The LCU scheme for $\hat{P}_N$ and $\hat{P}_{M_S}$ projectors, if number of LCU terms is power of two. The state $\hat{P}_{o_i}|\psi\rangle$ is up to corresponding normalization.}
    \label{fig:LCU_proj}
\end{figure}
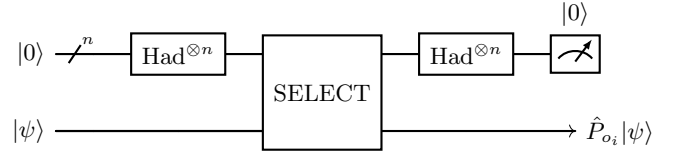

All weights in Eq.~\eqref{eq:finite_sum_proj} are equal. 
Consequently, the LCU ancilla preparation operator (PREP) reduces to Hadamards when the number of terms is a power of two.
The SELECT operator consists of controlled unitaries $e^{i \phi_k (o_i - \hat{O})}$ for different angles $\phi_k$.
The construction of the SELECT is provided in Appendix~\ref{appendix:primitives}.
The complete LCU circuits shown in Eq.~\eqref{eq:finite_sum_proj} with optimal PREP is shown in Fig.~\ref{fig:LCU_proj}. 
The measurement and postselection of the ancilla register in the $|0\rangle$ state allow for the implementation of projectors $\hat{P}_N$ and $\hat{P}_{M_S}$.

\noindent\textbf{$\hat{S}^2$ operator.}

The total-spin projector $\hat{P}_{S,M_S}$ can be implemented using numerical discretization of the group integral in Eq.~\eqref{eq:proj_int_S2}.  
We discretize the $SU(2)$ projector for a fixed $(S,M_S)$ sector using Gauss--Legendre nodes at the polar angle and uniform trapezoid grids at the azimuthal angles as follows: 
\begin{equation}
\label{eq:S2_discrete}
\hat P_{S,M_S}\;\approx\;
\sum_{b=1}^{N_\beta}\sum_{a=0}^{N_\alpha-1}\sum_{g=0}^{N_\gamma-1}
w_{bag}\; R(\alpha_a,\beta_b,\gamma_g),
\end{equation}
where the weights are defined as:
\begin{equation}
\label{eq:weights_matched}
\begin{aligned}
&w_{bag}\;=\;\frac{2S+1}{2N_\alpha N_\gamma}\;w_b^{\rm GL}\;D^{S}_{M_SM_S}(\alpha_a, \beta_b, \gamma_g)^{*},\\
&\beta_b = \arccos (x_b).    
\end{aligned}
\end{equation}
Here, $x_b$ and $w_b^{\rm GL}$ are the nodes and weights for the Gauss–Legendre quadrature with $N_\beta$ nodes.
In the implementation, the LCU uses $|w_{bag}|$ as the amplitudes, whereas the phase is absorbed into $R(\alpha_a,\beta_b,\gamma_g)$ using an additional phase gate.

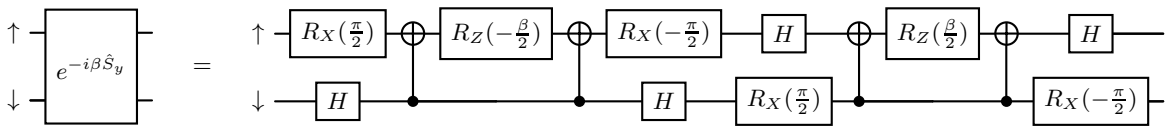
\begin{figure*}[t]
\centering
\begin{quantikz}[column sep=0.2cm,row sep={0.9cm,between origins}]
        \lstick{$\uparrow$} & \gate[2]{e^{-i \beta \hat{S}_y}} &\\ 
        \lstick{$\downarrow$} & \qw &
\end{quantikz}
    \quad $=$ \quad
\begin{quantikz}[column sep=0.2cm, row sep={0.9cm,between origins}]
\lstick{$\uparrow$} 
    & \gate{R_X(\tfrac{\pi}{2})}
    & \targ{} & \gate{R_Z(-\tfrac{\beta}{2})} & \targ{}
    & \gate{R_X(-\tfrac{\pi}{2})}
    & \gate{H} & \targ{} & \gate{R_Z(\tfrac{\beta}{2})} & \targ{} & \gate{H}
    & \qw \\
\lstick{$\downarrow$} 
    & \gate{H}
    & \ctrl{-1} & \qw & \ctrl{-1} 
    & \gate{H}
    & \gate{R_X(\tfrac{\pi}{2})} & \ctrl{-1} & \qw & \ctrl{-1} & \gate{R_X(-\tfrac{\pi}{2})}
    & \qw
\end{quantikz}
\caption{Circuit representation of $e^{-i \beta S_y}$ for a single spin-orbital pair $(\uparrow,\downarrow)$. This fragment is repeated for all spatial orbitals.}
\label{fig:ExpSy}
\end{figure*}

Each $R(\alpha,\beta,\gamma) = e^{-i\alpha \hat{S}_z} e^{-i\beta \hat{S}_y} e^{-i\gamma \hat{S}_z}$ consists of two $\hat{S}_z$ rotations, whose circuit is illustrated in Fig.~\ref{fig:expSz}.
The $\hat{S}_y$ rotation can also be constructed efficiently but requires CNOT gates~\cite{Tsuchimochi2020}.
Figure~\ref{fig:ExpSy} shows a single two-spin-orbital fragment, which is repeated for each spin-orbital pair.
The complete circuit for $R(\alpha,\beta,\gamma)$ is the consecutive application of the $\hat{S}_z$, $\hat{S}_y$ and $\hat{S}_z$ rotations.

The overall projectors in Eq.~\eqref{eq:S2_discrete} is then realized as a weighted LCU over these rotation unitaries, requiring the ancilla register, consisting of $\lceil\log_2 (N_\alpha N_\beta N_\gamma)\rceil$ qubits.
The PREP operator corresponding to $\beta$ angles is nontrivial and cannot be replaced by Hadamard gates.

As mentioned in Sec.~\ref{subsec:symmetry_projectors}, the qubit cost can be reduced if the initial guess is of definite $\hat{S}_z$ symmetry.
In this case, only a sequential application of $\hat{P}_S$ and $\hat{P}_{M_S}$ projectors is required. 
Then, the number of ancillary qubits is $\lceil\log_2 (N_\alpha N_\beta)\rceil$.
The gate cost is estimated in Sec.~\ref{sec:results}.

\subsubsection{Generalized quantum signal processing}
\label{subsubsec:GQSP}

Both the $\hat{N}$ and $\hat{S}_z$ operators have uniformly spaced eigenvalues that allow their projectors to be expressed as polynomials of the corresponding unitary operator. 
Formally, the projector in Eq.~\eqref{eq:finite_sum_proj} can be rewritten as a polynomial
\begin{equation}
    \hat{P}_{o_i} = \operatorname{Poly}(\hat{U}) = \sum_{k=0}^{d} c_k \hat{U}^k,
    \label{eq:GQSP_proj}
\end{equation}
where $\hat{U} = e^{i \frac{2\pi}{N_\phi} \hat{O}}$ is the signal unitary and $c_k = \frac{1}{N_\phi} e^{i \frac{2\pi k}{N_\phi} o_i}$ are complex polynomial coefficients.

The resulting polynomial transformation can be implemented using GQSP with only a single ancilla qubit.
GQSP is essential because the polynomial $\operatorname{Poly}(\hat{U})$ is of mixed parity and has complex coefficients.
The degree of the polynomial $d$ required for exact projection is equal to the number of spin-orbitals $N_\text{SO}$. 
The general structure of a GQSP circuit that implements such a projector is explained in Appendix~\ref{appendix:primitives}.

A detailed comparison of qubit requirements and circuit depths for LCU and GQSP implementations is presented in Sec.~\ref{sec:results}.

\subsubsection{Generalized quantum singular value transformation}

Another approach to constructing symmetry projectors relies on the Lagrange interpolation formula introduced in Sec.~\ref{subsec:symmetry_projectors}. 
In this method, the projector is expressed as a polynomial with the corresponding symmetry operator. 
\begin{equation}
    \hat{P}_{o_i} = \operatorname{Poly}(\hat{O}) = \sum_{k=0}^{d} p_k \hat{O}^k,
\end{equation}
where $p_k$ denotes the polynomial coefficients, defined in Eq.~\ref{eq:lagrange_proj}.
The major limitation of the Lagrange interpolation approach, when implemented directly, is the need to compute the high powers of $\hat{O}$, which results in an exponentially growing number of Pauli terms, and therefore, quickly becomes impractical. 
The GQSVT solves this issue by allowing for the efficient implementation of polynomial transformations of block-encoded operators.

The difference from the previous approach is that we use LCU not to encode the projector operators ($\hat{P}_N$, $\hat{P}_{M_S}$, $\hat{P}_{S}$), but the symmetry operators ($\hat{N}$, $\hat{S}_z$, and $\hat{S}^2$) using Eqs.~\eqref{eq:N_op_JW}, \eqref{eq:Sz_op_JW} and \eqref{eq:S2_op_JW}. 
Subsequently, for each operator, we construct a corresponding polynomial that implements a Löwdin-like projector, which selectively amplifies the desired eigenvalue sector. 

Considering that the symmetry operators are Hermitian and therefore diagonalizable, the singular value transformation is equivalent to a polynomial transformation for the symmetry operators.
The GQSVT protocol is then applied to the block-encoding operators to realize the projector operator exactly.
The degree of the polynomial for $\hat{P}_N$ and $\hat{P}_{M_S}$ projectors is equal to $N_\text{SO}$, whereas for $\hat{P}_{S^2}$ it is shorter and equal to the number of distinct eigenvalues for the $\hat{S}^2$ operator $N_\text{SO}/2$.
The degree of the implemented polynomial determined the circuit depth.
This construction uses $1$ signal ancilla and $\lceil\log_2L\rceil$ ancilla qubits, one for the signal register and additional ancillas for LCU block encoding. 
The general architecture of the GQSVT projector circuit is shown in the Appendix~\ref{appendix:primitives}.

The primary advantage of this approach is its generality: the same framework can be used to construct projectors for arbitrary symmetry operators without modification. 
Furthermore, multiple eigenvalue sectors can be targeted simultaneously by appropriately designing a polynomial $\operatorname{Poly}(\hat{O})$. 
Unlike the numerically discretized integral constructions used for $\hat{P}_{S,M_S}$, the GQSVT-based approach provides an exact polynomial realization and therefore offers higher theoretical accuracy.
However, the Löwdin projector for $\hat{S}^2$ does not project onto a particular $M_S$ sector.

Nevertheless, the practical implementation of GQSVT-based projectors remains challenging owing to the circuit depth and required gate precision. 
These aspects are analyzed in detail in Sec.~\ref{sec:results}, where we compare the resource requirements and discuss their potential applications as state-preparation filters for QPE.

%% file: Sections/03_Computational_details.tex
\section{Computational details} 
\label{sec:computation}

The software stack used in this study is summarized below.

Resource estimation and fault–tolerant circuit compilation were performed using the \textsc{QURI Parts} framework (\textsc{QSub} subpackage)~\cite{quri_parts}. 
Numerical circuit simulations were performed using \textsc{Qulacs}~\cite{qulacs}.
The second-quantized Hamiltonian and Jordan–Wigner mappings were generated using \textsc{OpenFermion}~\cite{openfermion}, and the molecular electronic integrals were obtained from \textsc{PySCF}~\cite{pyscf2018}.

All the scripts and notebooks used to produce the figures and tables in this study have been released at:
\url{https://github.com/mizukami-group/gqsp_symmetry_projectors}.

%% file: Sections/04_Results.tex
\section{Results and discussion} 
\label{sec:results}
\subsection{Theoretical scaling}
\label{subsec:theory}

We assessed the projector implementations in terms of the asymptotic gate counts and ancilla requirements. 
T gate counts are crucial in fault-tolerant settings because each T gate requires magic-state distillation. 
CNOT counts are also informative because they correlate with the circuit depth and routing overhead in limited-connectivity architectures where SWAP operations can introduce significant errors. 
Temporary work ancilla qubits that are uncomputed deterministically and can be reused freely are not included.

We begin our analysis with LCU based projectors for the $\hat{N}$ and $\hat{S}_z$ operators.
For this construction, the number of ancilla qubits required for indexing scales as $O(\log N_\text{SO})$.
The costs are determined by the PREP (ancilla state preparation) and SELECT (controlled operator selection) components.

When the number of LCU terms is a power of two, PREP reduces to a layer of Hadamard gates.
In the general case, PREP prepares a superposition of the $O(N_\text{SO})$ terms.
A binary-tree state preparation scheme yields $O(N_\text{SO} \log (N_\text{SO}/\epsilon))$ T and $O(N_\text{SO})$ CNOT gates.
Using QROM reduces the T count to $O(N_\text{SO})$ at the expense of $N_\text{SO} \log (N_\text{SO}/\epsilon)$ CNOT gates and a modest increase in ancilla qubits by $O(\log (1/\epsilon))$ qubits, depending on the required PREP precision $\epsilon$.

SELECT uses indexed single-qubit unitaries.
Considering the unitary constructions shown in Fig.~\ref{fig:expN} and Fig.~\ref{fig:expSz}, it is necessary to apply $N_\text{SO}$ single qubit rotations per LCU term.
The number of LCU terms is $O(N_\text{SO})$.
Therefore, any implementation of SELECT that uses indexed operations requires $O(N^2_\text{SO})$ CNOT gates.
The T gate cost includes both the indexing overhead and the synthesis of many small-angle rotations, $O(N^2_\text{SO}) + N^2_\text{SO}T(R_z)$, where $T(R_z)$ is the T gate scaling for a single rotational gate.
Corresponding to the Solovay–Kitaev theorem, the rotation gates require $O(\log (N_\text{SO}/\epsilon))$ T gates for a global precision budget $\epsilon$, which yields a total T gate count of $O(N^2_\text{SO} \log (N_\text{SO}/\epsilon))$.

Thus, for both $\hat{P}_N$ and $\hat{P}_{M_S}$, the cost is dominated by $O(N^2_\text{SO})$ controlled rotations and does not depend heavily on the specific PREP/SELECT implementation.

For the $\hat{S}^2$ projector, we begin with the general implementation expressed in Eq.~\eqref{eq:S2_discrete}, which requires $\lceil\log(N_\alpha N_\beta N_\gamma)\rceil$ ancilla qubits.
$N_\alpha$ and $N_\gamma$ scale as $N_\text{SO}$, as in the case of $\hat{P}_N$ and $\hat{P}_{M_S}$.
$N_\beta$ can be considered independent of $N_\text{SO}$ and dependent only on the chosen spin $S$, which will be explained in Sec.~\ref{subsec:accuracy}.
Therefore, the ancillary qubit count is $O(\log N_\text{SO})$, which is the same as that for $\hat{N}$ and $\hat{S}_z$ but with a prefactor $2$.

For a general $\hat{P}_{S,M_S}$ projector, $O(N_\text{SO}^2)$ LCU terms exist, each of which consists of $O(N_\text{SO})$ of unitaries shown in Fig.~\ref{fig:ExpSy} and are surrounded by the single qubit rotations shown in Fig.~\ref{fig:expSz}.
Thus, analogous to the previous cases of $\hat{P}_N$ and $\hat{P}_{M_S}$ projectors, the LCU is dominated by the number of unitary operators to be applied, which scales as $O(N_\text{SO}^3)$.
The resulting gate costs are $O(N_\text{SO}^3)$ CNOT gates and $O(N^3_\text{SO} \log (N_\text{SO}/\epsilon))$ T gates.
However, by expressing $\hat{P}_{S,M_S}$ as the product of $\hat{P}_{M_S}\hat{P}_{S}\hat{P}_{M_S}$ (see Eq.~\eqref{eq:proj_int_S2_PPP}) significantly reduced the scaling to that of a single $\hat{P}_{M_S}$ projector.
If the input is already an $\hat{S}_z$ eigenstate, one of the $\hat{P}_{M_S}$ projectors can be omitted without affecting the asymptotic scaling.

\begin{table*}[!tbp]
\centering
\caption{Asymptotic scaling of T and CNOT gate counts and number of ancilla qubits for different projector implementations. $N_\text{SO}$ is the number of spin-orbitals and $\epsilon$ is the target projector precision.
$(\hat{S}^2, \hat{S}_z)$ indicates a simultaneous projector onto total spin $S$ and its $z$-component $M_S$.}
\label{tab:gate_scalings}
\begin{tabular}{lccccc}
\hline\hline
\textbf{Method} & \textbf{Projector type} & \textbf{Implementation} & \textbf{T gates} & \textbf{CNOT gates} & \textbf{Ancillas} \\
\hline
LCU & $\hat{N}, \hat{S}_z$ & Any LCU & $O(N_\text{SO}^2 \log (N_\text{SO}/\epsilon))$ & $O(N_\text{SO}^2)$ & $O(\log N_\text{SO})$ \\
LCU & $(\hat{S}^2, \hat{S}_z)$ & Any LCU/$\hat{P}_{S,M_S}$ (Eq.~\eqref{eq:proj_int_S2}) & $O(N^3_\text{SO} \log (N_\text{SO}/\epsilon))$ & $O(N^3_\text{SO})$ & $O(\log N_\text{SO})$ \\[4pt]
LCU & $(\hat{S}^2, \hat{S}_z)$ & Any LCU/$\hat{P}_{M_S}\hat{P}_{S}\hat{P}_{M_S}$(Eq.~\eqref{eq:proj_int_S2_PPP}) & $O(N^2_\text{SO} \log (N_\text{SO}/\epsilon))$ & $O(N^2_\text{SO})$ & $O(\log N_\text{SO})$ \\[4pt]
GQSP & $\hat{N}, \hat{S}_z$ & --- & $O(N_\text{SO}^2 \log (N_\text{SO}/\epsilon))$ & $O(N_\text{SO}^2)$ & $1$ \\[4pt]
GQSVT & $\hat{N}, \hat{S}_z$ & Naïve LCU & $O(N_\text{SO}^2 \log (N_\text{SO}/\epsilon))$ & $O(N_\text{SO}^2 \log N_\text{SO})$ & $O(\log N_\text{SO})$ \\
GQSVT & $\hat{N}, \hat{S}_z$ & QROM LCU & $O(N_\text{SO}^2)$ & $O(N_\text{SO}^2 \log (N_\text{SO}/\epsilon))$ & $O(\log (N_\text{SO}/\epsilon))$ \\[4pt]
GQSVT & $\hat{S}^2$ & Naïve LCU & $O(N_\text{SO}^3 \log (N_\text{SO}/\epsilon))$ & $O(N_\text{SO}^3 \log N_\text{SO})$ & $O(\log N_\text{SO})$ \\
GQSVT & $\hat{S}^2$ & QROM LCU & $O(N_\text{SO}^3)$ & $O(N_\text{SO}^3 \log (N_\text{SO}/\epsilon))$ & $O(\log (N_\text{SO}/\epsilon))$ \\
\hline\hline
\end{tabular}
\end{table*}

Consider now GQSP implementation for $\hat{N}$ and $\hat{S}_z$.
The polynomial degree $d$ in GQSP scales linearly with $O(N_\text{SO})$, as shown in Sec.~\ref{subsubsec:GQSP}.
The circuit requires $d$ of $SU(2)$ rotation gates for signal processing and $(d-1) N_\text{SO}$ controlled single qubit rotations to build the signal operator, yielding $O(N_\text{SO}^2)$ CNOT gates and $O(N^2_\text{SO} \log (N_\text{SO}/\epsilon))$ T gates.
This matches the asymptotic cost of LCU, but requires only one ancilla qubit.
The prefactors and robustness to angle precision differed and were evaluated separately.

GQSVT uses qubitization, which requires $O(\log L)$ ancilla qubits for $L$ LCU terms.
For $\hat{N}$ and $\hat{S}_z$ $L$ scales linearly $O(N_\text{SO})$; however, for $\hat{S}^2$, $O(N_\text{SO}^2)$.
This implies the same scaling of the ancilla qubits but a larger prefactor for $P_{S^2}$.
The degree of the interpolation polynomial is equal to the number of distinct eigenvalues minus one: $d = N_\text{SO}$ for $\hat{N}$ and $\hat{S}_z$ and $d = N_\text{SO}/2$ for $\hat{S}^2$.
The degree defines the number of times a qubitization operator is accessed.

For $\hat{N}$ and $\hat{S}_z$, the operators are decomposed into single-Pauli strings (see Eq.~\eqref{eq:N_op_JW} and Eq.~\eqref{eq:Sz_op_JW}).
Using the binary-tree PREP and indexed SELECT, we obtain $O(N^2_\text{SO} \log N_\text{SO})$ CNOT gates and $O(N^2_\text{SO} \log (N_\text{SO}/\epsilon))$ T gates.
QROM trades these counts to $O(N^2_\text{SO} \log (N_\text{SO}/\epsilon))$ CNOT gates and $O(N^2_\text{SO})$ T gates with ancilla $O(\log(N_\text{SO}/\epsilon))$.

For $\hat{S}^2$, the longest Pauli strings have constant length (up to $4$); however, the number of LCU terms is $O(N_\text{SO}^2)$ (see Eq.~\eqref{eq:S2_op_JW}).
This increases the cost by one power of $N_\text{SO}$, giving $O(N_\text{SO}^3 \log N_\text{SO})$ CNOT gates and $O(N_\text{SO}^3 \log (N_\text{SO}/\epsilon))$T gates for binary-tree implementation, and $O(N^3_\text{SO} \log (N_\text{SO}/\epsilon))$ CNOT gates and $O(N^3_\text{SO})$ T gates for QROM.
All scalings are summarized in Table~\ref{tab:gate_scalings}.

\subsection{Success probability}
\label{subsec:succ_prob}
The success probabilities for all these schemes are essentially the same in the best case and are equal to
\begin{equation}
\label{eq:succ_prob}
    p_\text{succ} = ||\hat{P}_{o_i} |\psi\rangle||_2^2 = |\langle\psi|\hat{P}_{o_i}|\psi\rangle|,
\end{equation}
where $\hat{P}_{o_i}$ is the projector onto the subspace, corresponding to $o_i$ and $|\psi\rangle$ is the initial wave function. 

This probability can be increased by using amplitude amplification (AA).
After $m$ rounds of amplification,
\begin{equation}
\label{eq:succ_prob_AA}
    p^\text{(m)}_\text{succ} = \sin^2((2m+1)\theta), \quad \theta = \arcsin(\sqrt{p_\text{succ}}).
\end{equation}

Choosing $m$ to maximize $p^\text{(m)}_\text{succ}$ yields:
\begin{equation}
\label{eq:AA_iters}
    m = \Big\lfloor\frac{\pi}{4 \theta} - \frac12\Big\rfloor \approx \frac{\pi}{4} \frac{1}{\sqrt{p_\text{succ}}},
\end{equation}
where $p_\text{succ}$ is assumed small.
Each round of amplification uses the underlying block a constant number of times; therefore, the gate count and depth grow linearly in $m$ (approximately by a factor of $2m+1$, plus the cost of reflections).

The number of iterations required to start from a uniform superposition can be estimated.
The uniform superposition state of all possible occupations of $N_\text{SO}$ spin-orbitals was projected onto a subspace with $N_\text{elec} = N_\text{SO}/2$ electrons.
Then
\begin{equation}
    |\langle\psi|\hat{P}_{N_\text{elec}}|\psi\rangle| = \frac{\binom{N_\text{SO}}{N_\text{elec}}}{2^{N_\text{SO}}} \sim \frac{1}{\sqrt{N_{SO}\pi/2}}.
\end{equation}
By substituting this expression into Eq.~\eqref{eq:succ_prob}, and then into Eq.~\eqref{eq:AA_iters}, we observe that the number of amplification steps increases only with the fourth root of $N_\text{SO}$, which is typically modest in practice.

\subsection{Numerical accuracy of projectors}
\label{subsec:accuracy}

To enable meaningful resource analysis, we quantified the numerical accuracy of the projectors and their dependence on discretization parameters. 
In the integral construction of Eqs.~\eqref{eq:proj_int} and \eqref{eq:proj_int_S2_PPP}, the only tunable parameter is the number of quadrature nodes.
This choice sets both the number of LCU terms and, in GQSP/QSVT-style realizations, the degree of the resulting polynomial.

\begin{figure*}
    \centering
    \includegraphics[width=1\linewidth]{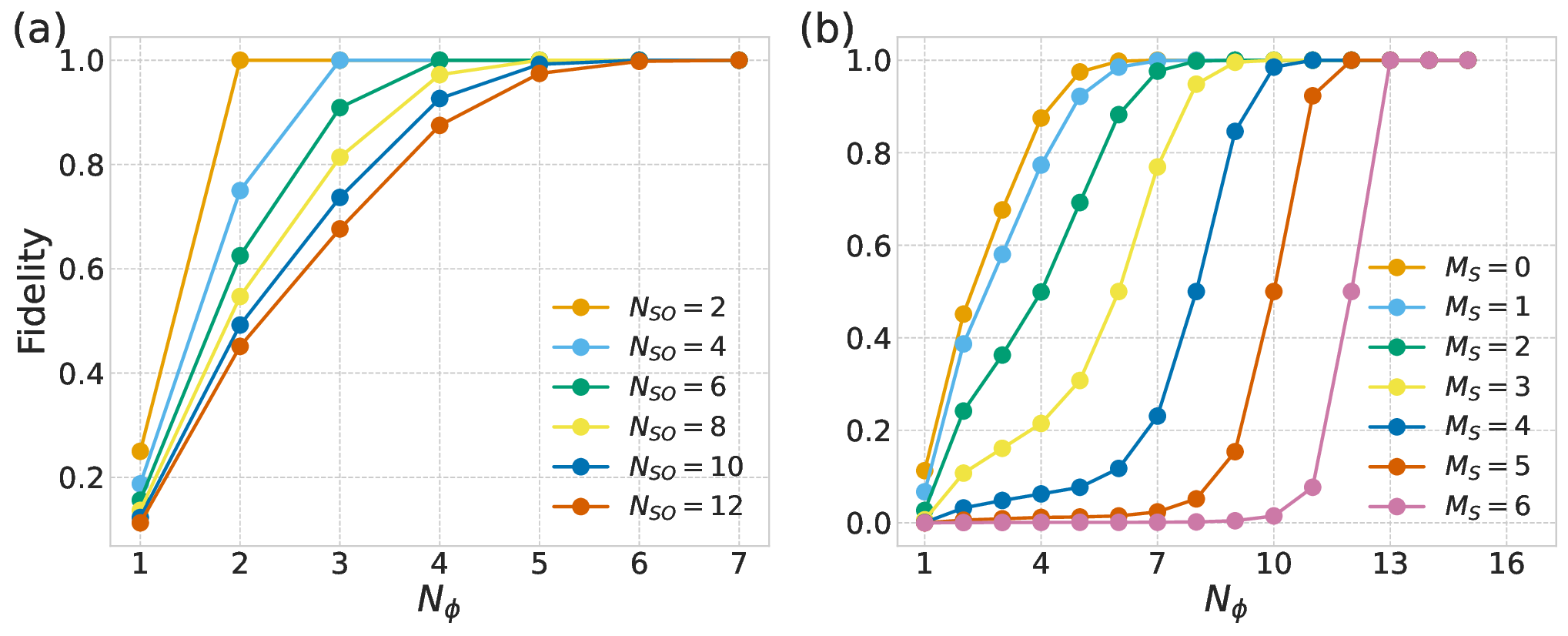}
    \caption{Fidelity of the $\hat{P}_{M_S}$ projector. (a) Dependence of the fidelity on the number of integration nodes for different numbers of spin-orbitals. (b) Dependence of the fidelity on the number of integration nodes for different target values of $M_S$.}
    \label{fig:num_acc_Sz}
\end{figure*}

\begin{figure}
    \centering
    \includegraphics[width=1\linewidth]{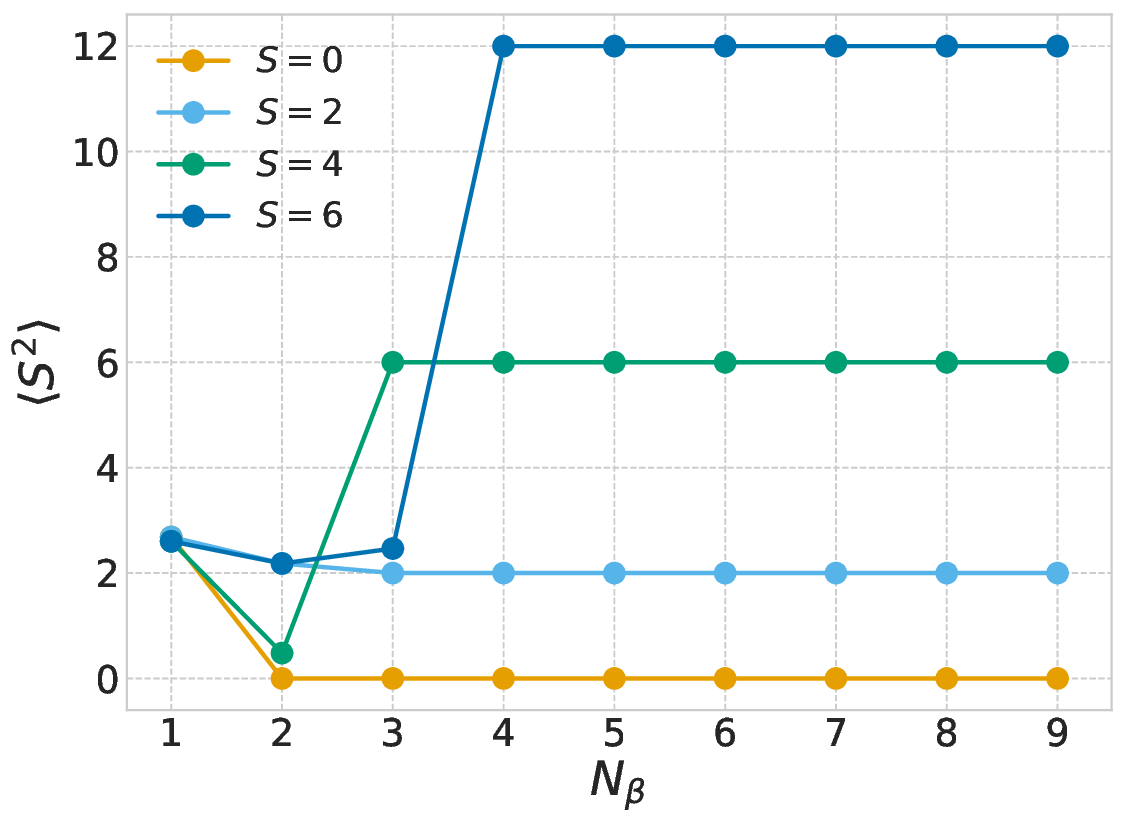}
    \caption{Expectation value of $\hat{S}^2$ for states produced by $\hat{P}_{S,M_S}$, as a function of the number of $\beta$-nodes $N_\beta$, for different target spin sectors $S$ at fixed $M_S=0$.}
    \label{fig:num_acc_S2}
\end{figure}

When the eigenvalues of a symmetry operator $\hat{O}$ are integer-spaced, the integral form of the corresponding projector can be evaluated exactly using uniform discretization as shown in Eq.~\eqref{eq:finite_sum_proj}. 
In the eigenbasis $\hat O|o_j\rangle=o_j|o_j\rangle$ the integrand in Eq.~\eqref{eq:proj_int} is reduced to a single Fourier mode
$e^{i\phi(o_i-o_j)}$ with integer frequency $q=o_i-o_j\in\mathbb Z$, so the continuous projector simply enforces
$\frac{1}{2\pi}\int_{0}^{2\pi} e^{iq\phi}d\phi=\delta_{q,0}$.
Replacing the integral with a uniform grid $\phi_k=2\pi k/N_\phi$ yields the corresponding discrete Fourier sum
$\frac{1}{N_\phi}\sum_{k=0}^{N_\phi-1} e^{i2\pi k q/N_\phi}$, which is exactly $1$ if $q\equiv 0\ (\mathrm{mod}\ N_\phi)$ and $0$ otherwise.
Therefore, the discrete projector is equal to the continuous projector if and only if no two distinct eigenvalues of $(o_i-\hat O)$ coincide with the modulo $N_\phi$; this is called the aliasing condition.

For $\hat{S}_z$, the spectrum in full Fock space is $\{-N_{\mathrm{SO}}/2,\ldots,N_{\mathrm{SO}}/2\}$ in unit steps. 
The accuracy of the discrete projector onto $M_S$ is guaranteed when
\begin{equation}
    N_\phi \geq \frac{N_{SO}}{2} + |M_S|+1.
\end{equation}
Therefore, in the common case $M_S=0$ the minimum exact choice is $N_\phi = N_{\mathrm{SO}}/2 + 1$. 
This reduces the prefactor for applying $\hat{P}_{M_S}$ relative to $N_\phi = N_{\mathrm{SO}} + 1$.

Figure~\ref{fig:num_acc_Sz} illustrates the fidelity of the $\hat{S}_z$ projector. 
We begin with uniform superposition. 
In panel (a), we fixed $M_S=0$ and varied $N_{\mathrm{SO}}$; in panel (b), we fixed $N_{\mathrm{SO}}=12$ and varied the target $M_S$.

For the particle-number operator $\hat{N}$, the analysis is analogous. 
With the eigenvalues $\{0,\ldots,N_{\mathrm{SO}}\}$, the aliasing condition for the projector onto $N_{\mathrm{elec}}$ is
\begin{equation}
    N_\phi \geq \frac{N_\text{SO}}{2} + \Bigg|N_\text{elec} - \frac{N_\text{SO}}{2}\Bigg| + 1.
\end{equation}
In particular, at half filling $N_{\mathrm{elec}}=N_{\mathrm{SO}}/2$, one can again take $N_\phi = N_{\mathrm{SO}}/2 + 1$.

The numerical integration of the total-spin projector $\hat{P}_{S,M_S}$ behaves differently. 
In our benchmarks, we target singlets with $M_S=0$, starting with a uniform superposition restricted to $M_S=0$. 
In this setting, it is convenient to apply $\hat{P}_S$ (see Fig.~\ref{fig:ExpSy}), followed by $\hat{P}_{M_S}$. 
For the $\hat{S}_z$ step, we use $N_\phi = N_{\mathrm{SO}} + 1$ nodes to make the discretization exact for all $M_S$, and for $M_S=0$, we use the minimal exact value $N_\phi=N_{\mathrm{SO}}/2+1$.

From Fig.~\ref{fig:num_acc_S2} we observe that the number of nodes $N_\beta$ for $\hat{P}_S$ required to saturate the correct projection grows with $S$.
In the range tested, we find $N_\beta=2$ for $S=0$, $N_\beta=3$ for $S=2$ and $S=4$, and $N_\beta=4$ for $S=6$.

This behavior can be understood by expanding the $\beta$-kernel in Eq.~\eqref{eq:proj_int_S2_PPP} in a basis with fixed $M_S$ and $S'$,
\begin{equation}
\begin{split}
&\langle S' M_S|e^{-i\beta S_y} | S' M_S\rangle\, d^S_{M_S M_S}(\beta)^* \\
&=d^{S'}_{M_S M_S}(\beta)\,d^S_{M_S M_S}(\beta)^* \\
&=\sum_{L=|S'-S|}^{S'+S} B_L(S,S',M_S)\,P_L(\cos\beta),
\end{split}
\end{equation}
where the coefficients $B_L(S,S',M_S)$ are independent of explicit $N_{\mathrm{SO}}$.
In particular, for the target component $S'=S$ the expansion contains only degrees up to $L_{\max}=2S$.

If one aims for a sufficient condition for an exact evaluation of the $S'=S$ contribution, the Gauss--Legendre quadrature with $N_\beta$ nodes integrates polynomials up to the degree $2N_\beta-1$, implying a conservative requirement $N_\beta\ge S+1$.
However, in our benchmarks, the expectation values were saturated for a significantly smaller $N_\beta$, as reported above.

In contrast, GQSVT implementations based on Lagrange interpolation do not allow for this freedom.
Their polynomial degree must be equal to the number of possible eigenvalues of the symmetry operator to filter out all the undesired eigenvalues.
Otherwise, if some eigenvalues are not filtered, the Lagrange polynomial can take large values in these sectors.
If the initial wave function is not perfectly zero in those sectors, it will be multiplied uncontrollably in those sectors, resulting in an even worse state than before the projection.
Another problem is that if the polynomial grows significantly in these sectors, it requires normalization proportional to the largest polynomial value.
This reduces the probability of success by the factor of the normalization constant.
Thus, the polynomial degree for GQSVT should always be $d = N_\text{SO} + 1$ for $\hat{N}$, $\hat{S}_z$ and $d = N_\text{SO}/2 + 1$ for $\hat{S}^2$.

\subsection{Gate precision sensitivity}
\label{subsec:precision}

The second practical issue for implementation and resource estimation is the robustness of the projectors to finite precision in single-qubit rotation gates.
This directly indicates the T gate counts required to synthesize these rotations and the resulting fidelity.
We modeled the finite precision by quantizing each continuous rotation angle $\phi$ to the nearest point on a uniform grid with a spacing $2\epsilon_r$.
Concretely, we replace
\begin{equation}
\label{eq:angle_quantization}
    \phi \;\mapsto\; \phi_d \equiv 2\epsilon_r \Big\lfloor \frac{\phi}{2\epsilon_r} + \frac12 \Big\rfloor
\end{equation}
that guarantees that the bound
\begin{equation}
\label{eq:angle_quantization_bound}
    |\phi-\phi_d| \le \epsilon_r .
\end{equation}
A smaller $\epsilon_r$ corresponds to a finer angular resolution, whereas a larger $\epsilon_r$ yields a coarser discretization.

\begin{figure}[t]
    \centering
    \includegraphics[width=1\linewidth]{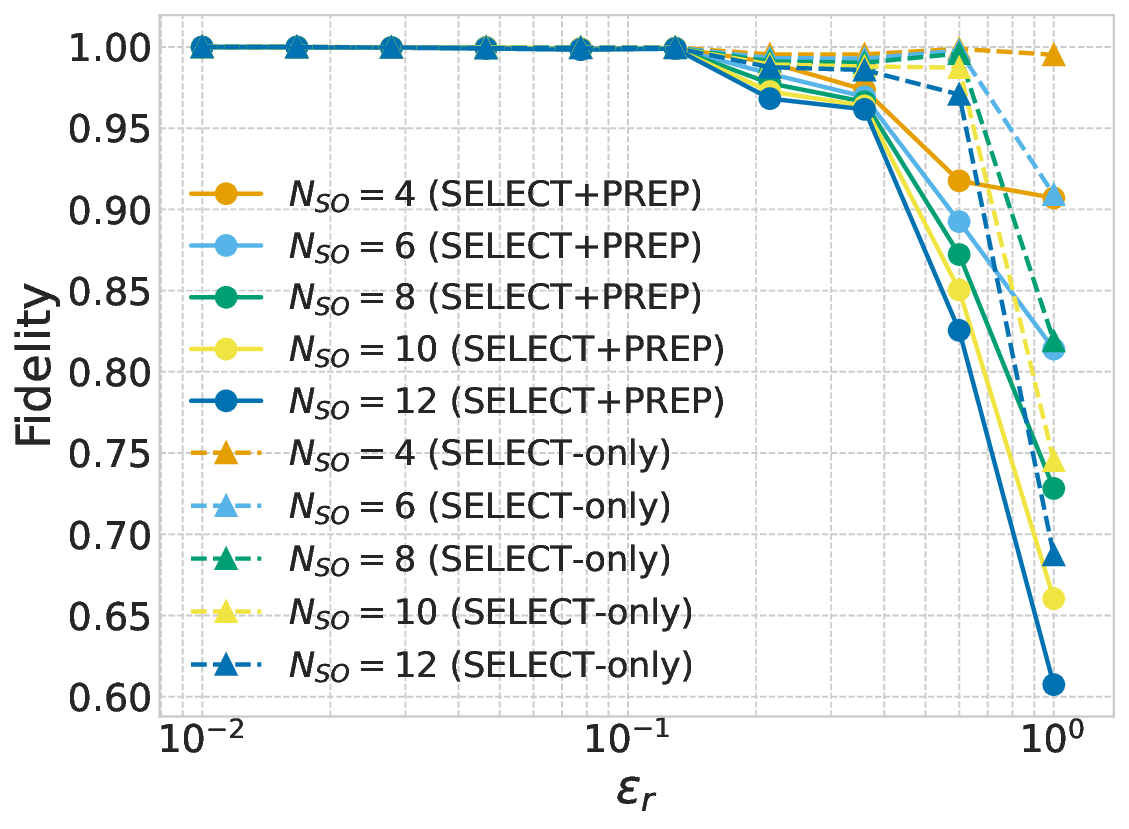}
    \caption{Fidelity of the LCU implementation of $\hat{P}_{M_S}$ versus single-qubit rotation precision $\epsilon_r$ for fixed $M_S=0$ and varying $N_{\mathrm{SO}}$. Solid lines with circular markers: both PREP and the indexed unitaries in SELECT are rounded (SELECT+PREP). Dashed lines with triangular markers: only the indexed unitaries in SELECT are rounded, while PREP is exact (SELECT-only).}
    \label{fig:angle_lcu_sz}
\end{figure}

We benchmarked the LCU implementation of $\hat{P}_{M_S}$ for singlet targets starting from a uniform superposition without fixing the particle number. 
Figure~\ref{fig:angle_lcu_sz} shows that the fidelity improves monotonically with angle precision and depends only weakly on the number of spin-orbitals $N_{\mathrm{SO}}$. 
We consider two models: (I) SELECT-only, where only the indexed unitaries are rounded and PREP is exact (model for QROM or trivial PREP at a power-of-two register size) and (II) SELECT+PREP, where both operators are rounded. 
PREP errors further reduced the fidelity, but not significantly. 
In all tested cases, $\epsilon_r=10^{-1}$ is sufficient to achieve high fidelity.

\begin{figure}[t]
    \centering
    \includegraphics[width=1\linewidth]{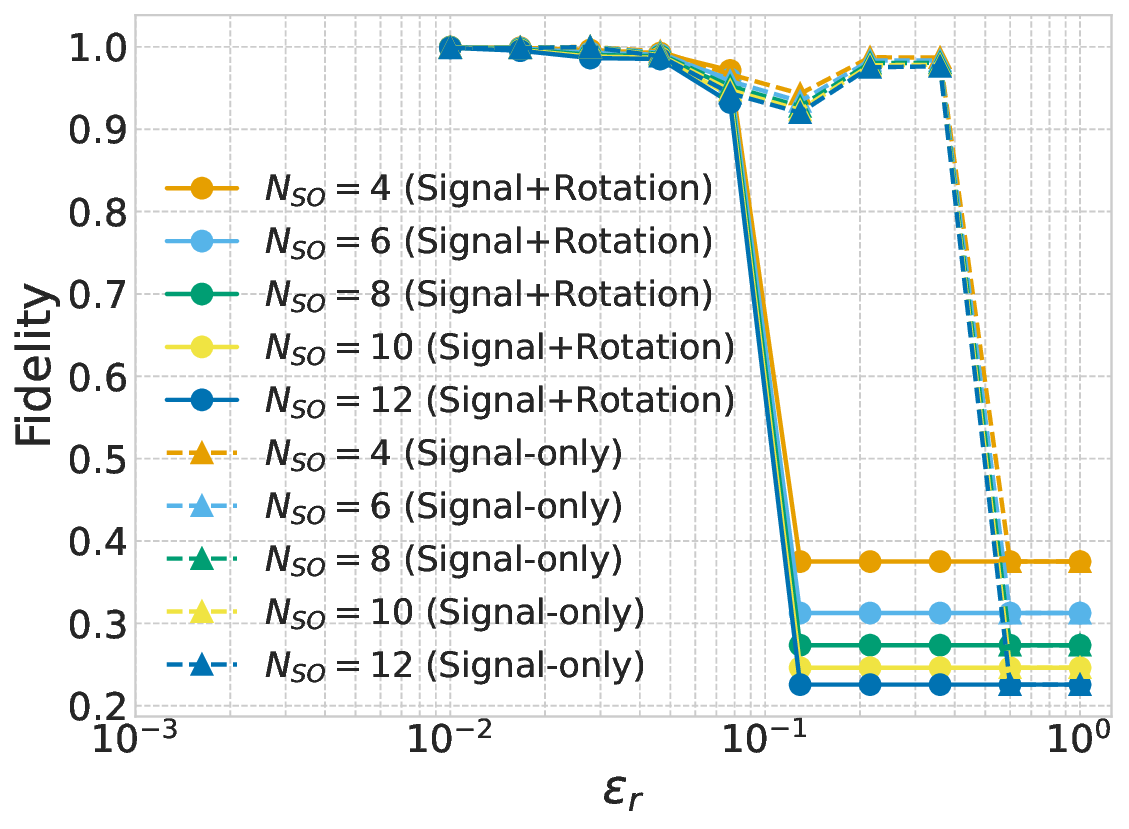}
    \caption{Fidelity of the GQSP implementation of $\hat{P}_{M_S}$ versus single-qubit rotation precision $\epsilon_r$ for fixed $M_S=0$ and varying $N_{\mathrm{SO}}$. Solid lines with circular markers: both the GQSP signal operator and the GQSP rotation sequence are rounded. Dashed lines with triangular markers: only the signal operator is rounded; the GQSP rotations are exact.}
    \label{fig:angle_gqsp_sz}
\end{figure}

Next, we tested the GQSP implementation for the same $\hat{P}_{M_S}$; the results are shown in Fig.~\ref{fig:angle_gqsp_sz}. 
The GQSP realization is more sensitive to the angle precision than LCU, although it is still relatively robust.
The signal operator angles were the most critical, and a precision of approximately $10^{-0.8}$ was generally considered sufficient.
The GQSP rotation sequence should be slightly more precise, better than $10^{-1.5}$, to avoid noticeable fidelity drops.
The dependence on $N_\text{SO}$ is small, and increasing the number of spin-orbitals does not significantly affect the required precision.
In all tests with $\epsilon_r = 10^{-2}$ the fidelities were close to one.

\begin{figure}
    \centering
    \includegraphics[width=1\linewidth]{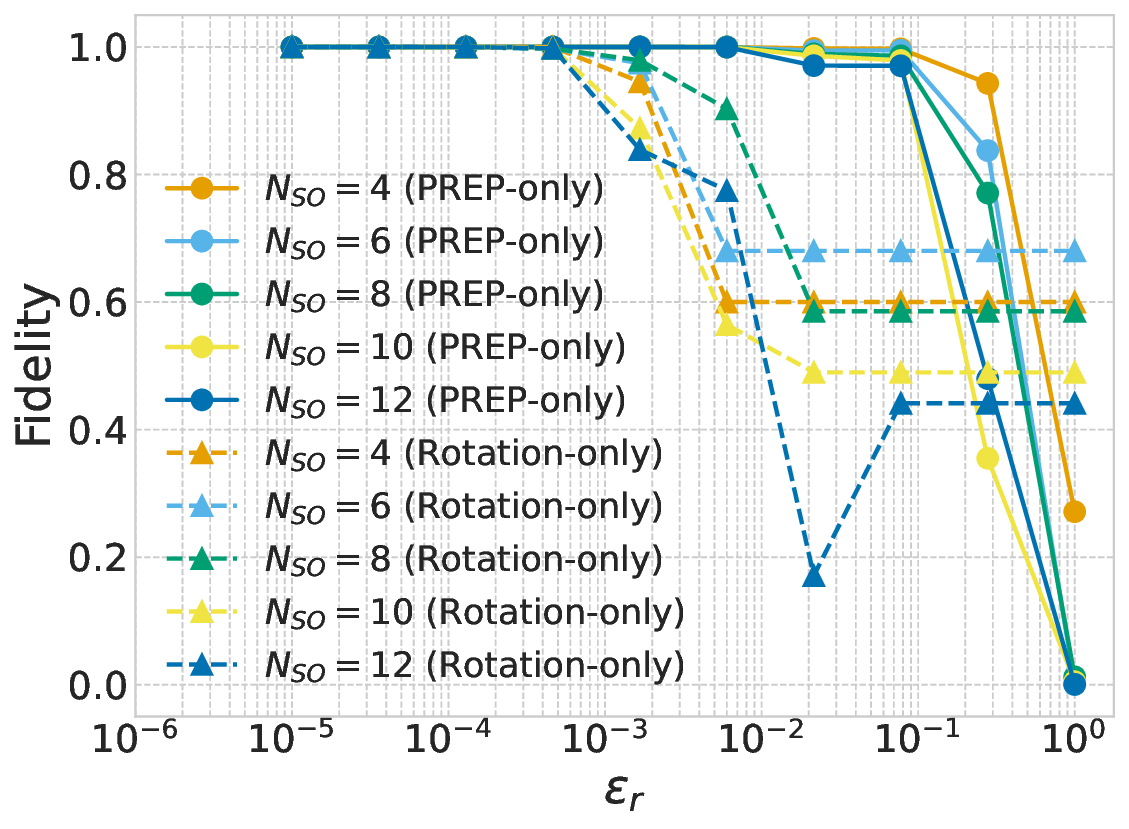}
    \caption{Fidelity of the GQSVT implementation of $\hat{P}_{M_S}$ versus single-qubit rotation precision $\epsilon_r$ for fixed $M_S=0$ and varying $N_{\mathrm{SO}}$. Solid lines with circular markers: GQSVT rotations are exact; PREP is rounded. Dashed lines with triangular markers: only GQSVT rotations are rounded; PREP is exact.}
    \label{fig:angle_gqsvt_sz}
\end{figure}

For the GQSVT implementation of $\hat{P}_{M_S}$ (see Fig.~\ref{fig:angle_gqsvt_sz}), the robustness was considerably lower.
We examined two precision models: one in which PREP was exact and GQSVT rotations were rounded, and the other in which the rotations were exact but PREP was rounded.
The results show that PREP can tolerate somewhat coarse precision, approximately $10^{-2}$, without a major loss of fidelity, but GQSVT rotations require much finer control.
A precision of at least $\epsilon_r\lesssim 10^{-3.7}$ is required to maintain a high fidelity.
Furthermore, accurate phase choices are critical for correctly realizing the intended polynomial transformation, and the sensitivity increases with $N_\text{SO}$, owing to the accumulation of small phase errors and numerical conditioning in the polynomial construction.

Because the standard Clifford+T synthesis has a cost that scales linearly with the number of target bits, reducing $\epsilon_r$ by one decimal order (adding $\log_2 10 \approx 3.3$ bits of precision) increases the per-rotation T gate count by a moderately constant factor. 
Thus, the qualitative thresholds above ($10^{-1}$ for LCU, $10^{-2}$ for GQSP, $10^{-4}$ for GQSVT) translate into predictable logarithmic increases in the T budget per rotation.

\subsection{Resource estimates}
\label{subsec:resources}

\begin{figure*}
    \centering
    \includegraphics[width=1\linewidth]{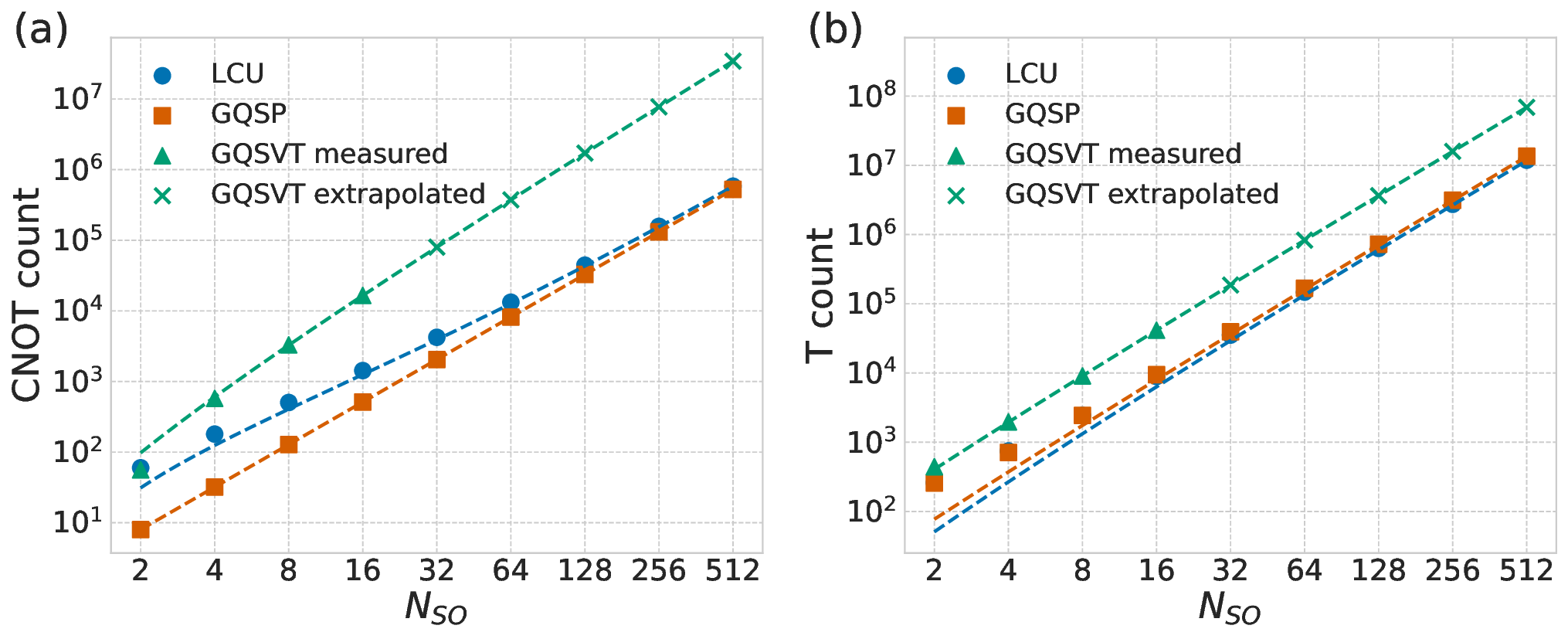}
    \caption{Gate counts for $\hat{P}_{M_S}$ as a function of $N_{\mathrm{SO}}$: (a) CNOTs and (b) $\mathrm{T}$ gates, comparing LCU, GQSP, and GQSVT implementations under the precision budgets summarized in the text.}
    \label{fig:S_z_gates_counts}
\end{figure*}

\begin{figure*}
    \centering
    \includegraphics[width=1\linewidth]{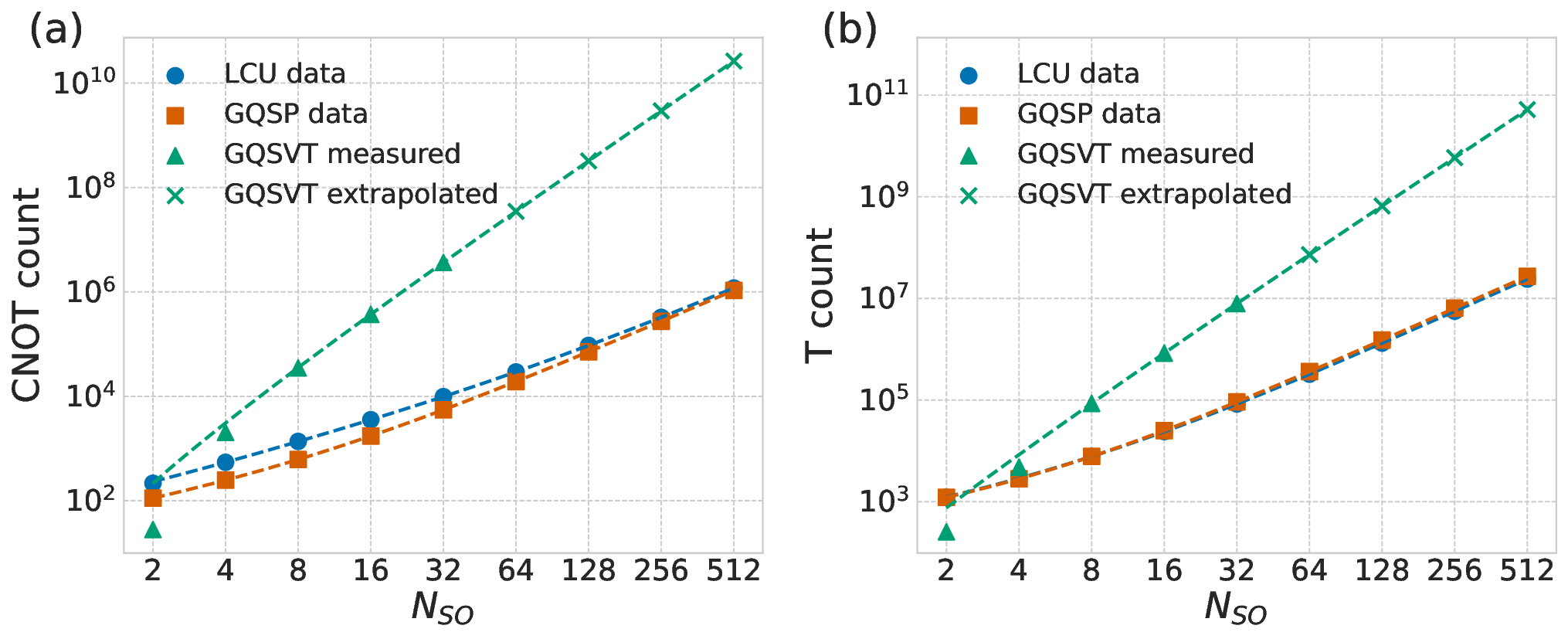}
    \caption{Gate counts for total-spin projectors versus $N_{\mathrm{SO}}$: (a) CNOTs and (b) $\mathrm{T}$ gates for $\hat{P}_{S,M_S}$ via LCU and GQSP, and for $\hat{P}_{S^2}$ via GQSVT.}
    \label{fig:S2_gates_counts}
\end{figure*}

We estimated the gate counts for the concrete implementations of the projectors. 
This allowed us to assess practicality. 
We counted CNOTs, Toffolis, and single-qubit rotations, then convert to T gate counts using the angle precisions established in the previous section.

We counted the Toffoli as $6$ CNOTs and $7$ T gates, which is the standard decomposition~\cite{nielsen_chuang_2000}. 
A single-axis rotation to precision $\epsilon_r$ costs
\begin{equation}
  T_{\rm rot}(\epsilon_r)\;=\;3\log_2(1/\epsilon_r)\;+\;o(\log(1/\epsilon_r))
\end{equation}
T gates using Clifford+T synthesis~\cite{morisaki2025optimalancillafreecliffordtsynthesis}. 
When $N_{\rm rot}$ independent rotations appear, we allocate a per-rotation budget $\epsilon_r/N_{\rm rot}$ such that the total error is at most $\epsilon_r$. 
This provides a conservative aggregate estimate, as follows:
\begin{equation}
  T_{\rm total\;rot}\;\approx\;3\,N_{\rm rot}\,\log_2\!\big(N_{\rm rot}/\epsilon_r\big).
\end{equation}

First, we compared the T and CNOT gate counts for the $\hat{P}_{M_S}$ projector (see Fig.~\ref{fig:S_z_gates_counts}).
The corresponding resource estimations for the particle number projector $\hat{P}_N$ are essentially the same because both rely on the same discretization structure.
In all the cases, we set the number of quadrature nodes to $N_\phi=N_{\mathrm{SO}}+1$, which guarantees the exactness of any desired value of $M_S$.
For the LCU construction, we choose the number of LCU terms to be a power of two; thus, the PREP blocks consist only of Hadamard gates.

In Fig.~\ref{fig:S_z_gates_counts}(a), we observe that both the LCU and GQSP realizations of $\hat{P}_{M_S}$ exhibit quadratic scaling in the number of CNOT gates.
In both cases, the dominant contribution increases as $2N_\text{SO}$, with an additional overhead for LCU which increases approximately as $N_\text{SO} \log N_\text{SO}$ owing to the use of indexed unitaries in the SELECT step.
Therefore, the LCU consistently requires more CNOT gates than GQSP for the same system size.
However, the difference between the two becomes less pronounced as the number of spin-orbitals increases because the shared quadratic term dominates the scaling.

In contrast, the GQSVT construction exhibited worser scaling.
For GQSVT, we used the standard binary-tree version of LCU.
The number of CNOT gates increases as $N_\text{SO}^2 \log N_\text{SO}$.
We were able to compute the exact counts up to $N_\text{SO} = 16$, and the same trend was easily extrapolated to larger systems up to at least $N_\text{SO} = 512$.
This growth makes GQSVT more expensive than LCU or GQSP in terms of entangling operations.

To estimate the number of T gates, we used the values from the precision study in Sec.~\ref{subsec:precision}.
This behavior was noticeably more uniform among the three methods (see Fig.~\ref{fig:S_z_gates_counts}(b)).
In all cases, we observed scaling proportional to $N_\text{SO}^2 \log N_\text{SO}$, reflecting the combined effect of the number of required single-qubit rotations and the synthesis cost required to reach the target precision.
Both LCU and GQSP implementations share essentially the same leading prefactor in this scaling; the numerical data indicate a coefficient of approximately $4$.
The two approaches differ only in additive lower-order terms; therefore, while LCU consistently yields a slightly lower T gate count than GQSP, the difference remains small across the entire range of problem sizes investigated.

Although the T gate scaling of GQSVT with the binary-tree LCU implementation has the same asymptotic form, the corresponding prefactor is significantly larger, approximately $20$ in our tests. 
This means that at the same target fidelity, GQSVT is approximately $5$ times more expensive in terms of T gates than LCU or GQSP. 
Thus, although all three methods share similar asymptotic scaling, GQSVT has a substantially higher constant cost and is the least practical option.

For the LCU and GQSP implementation of $\hat{P}_{S,M_S}$ projector, we use standard decomposition as $\hat{P}_{M_S} \hat{P}_{S} \hat{P}_{M_S}$; thus, the resource cost is determined by the implementation of $\hat{P}_{M_S}$.
Operators $\hat{P}_{S}$ were implemented using the LCU construction in both cases.
Therefore, the difference between LCU and GQSP arises only in the implementation of $\hat{P}_{M_S}$.
As before, we chose the number of quadrature nodes for $\hat{P}_{M_S}$ as $N_\phi = N_\text{SO} + 1$ to ensure the exactness of any spin projection.
For $\hat{P}_S$ because we focus on the singlet states ($S=0$), only two nodes are required in the integration; therefore, the cost of this step is independent of $N_\text{SO}$.

The GQSVT approach is slightly different in that it constructs the projector onto the desired total-spin subspace via $\hat{P}_{S^2}$ rather than by selecting a specific $M_S$.
However, in practice, this still produces states with a well-defined total spin suitable for the same applications; therefore, we include it in the comparison.

As shown in Fig.~\ref{fig:S2_gates_counts}(a), for both the LCU-based and GQSP-based implementations, the CNOT count scales quadratically with the number of spin-orbitals.
The leading contribution grows as $4N_\text{SO}$, with a slightly larger additive term for LCU because of the additional SELECT overhead inherited from $\hat{P}_{M_S}$.
However, this difference becomes smaller in comparison with the total gate count as $N_\text{SO}$ increases; hence, for moderate or large system sizes, the two methods are essentially comparable in terms of CNOT cost.

By contrast, the GQSVT implementation with the binary-tree LCU exhibited a much steeper scaling.
In this case, the CNOT count increases approximately as $21 N_\text{SO}^3 \log N_\text{SO}$, reflecting both a higher-degree sequence length and a more complex signal processing structure.
Consequently, GQSVT becomes significantly more expensive than LCU or GQSP when $N_\text{SO}$ has more than a few orbitals.
Nevertheless, for very small system sizes, the situation can be reversed; for example, at $N_\text{SO}=2$, the GQSVT implementation requires fewer CNOTs than LCU or GQSP.

The trends in the T gate counts are shown in Fig.~\ref{fig:S2_gates_counts}(b) exhibit similar patterns.
For both LCU and GQSP realizations, the T gate cost increases as $8 N_\text{SO}^2 \log N_\text{SO}$.
In direct comparison, GQSP grows slightly faster because of its larger rotation overhead; thus, the LCU construction retains a small advantage across all the studied system sizes.

For GQSVT with the binary-tree LCU, the T gate count grows substantially more quickly, as $36.5 N_\text{SO}^3 \log N_\text{SO}$, in line with the deeper signal-processing sequence and the increased synthesis precision required for stable polynomial filtering.
Consequently, GQSVT becomes significantly more expensive than both LCU and GQSP as the system size increases.
Similar to the CNOT comparison, a narrow regime exists at a very small $N_\text{SO}$, in which GQSVT can be competitive or even favorable; however, beyond this range, cubic scaling dominates.

\subsection{Projectors as a QPE filter}
\label{subsec:QPE_filter}

The success probability of QPE depends directly on the overlap between the prepared input state $|\psi\rangle$ and the target eigenstate $|\phi_0\rangle$ of the Hamiltonian. 
In particular, the probability of obtaining the correct eigenvalue is $|\langle\phi_0|\psi\rangle|^2$.
In the case of strongly correlated electronic states or when several eigenstates lie within a narrow energy window, preparing an input state with a sufficiently high overlap can be challenging. 
This issue is particularly pronounced for open-shell molecules and other systems with near-degeneracies.


To address this limitation, we apply symmetry projectors $\hat{P}$ prior to QPE. 
The operation $\hat{P}|\psi\rangle$ increases the weight of the components of $|\psi\rangle$ that are in the symmetry sector of the target eigenstate, while suppressing the components belonging to other sectors. 
Because $\hat{P}|\phi_0\rangle = |\phi_0\rangle$ for the eigenstates, this filtering step does not perturb the target state. 

We tested the effectiveness of projector filtering in systems in which the electronic spectrum exhibited near-degeneracy.

\textbf{O$_2$ Singlet–Triplet Gap.}
Molecular oxygen (O$_2$) is a canonical example in which the triplet ground state $X^3\Sigma_g^-$ is close to the low-lying singlets $a^1\Delta_g$ and $b^1\Sigma_g^+$. 
We consider an active space calculation in the CAS($8$,$6$) space within an aug-cc-pVTZ basis, where the orbitals are optimized in a triplet CASSCF calculation.
Active orbitals consisted of bonding and antibonding $\sigma$ pair and bonding and antibonding $2$ $\pi$ pairs; the O–O bond length was set to $1.208$ \AA.

\begin{figure}
    \centering
    \includegraphics[width=1\linewidth]{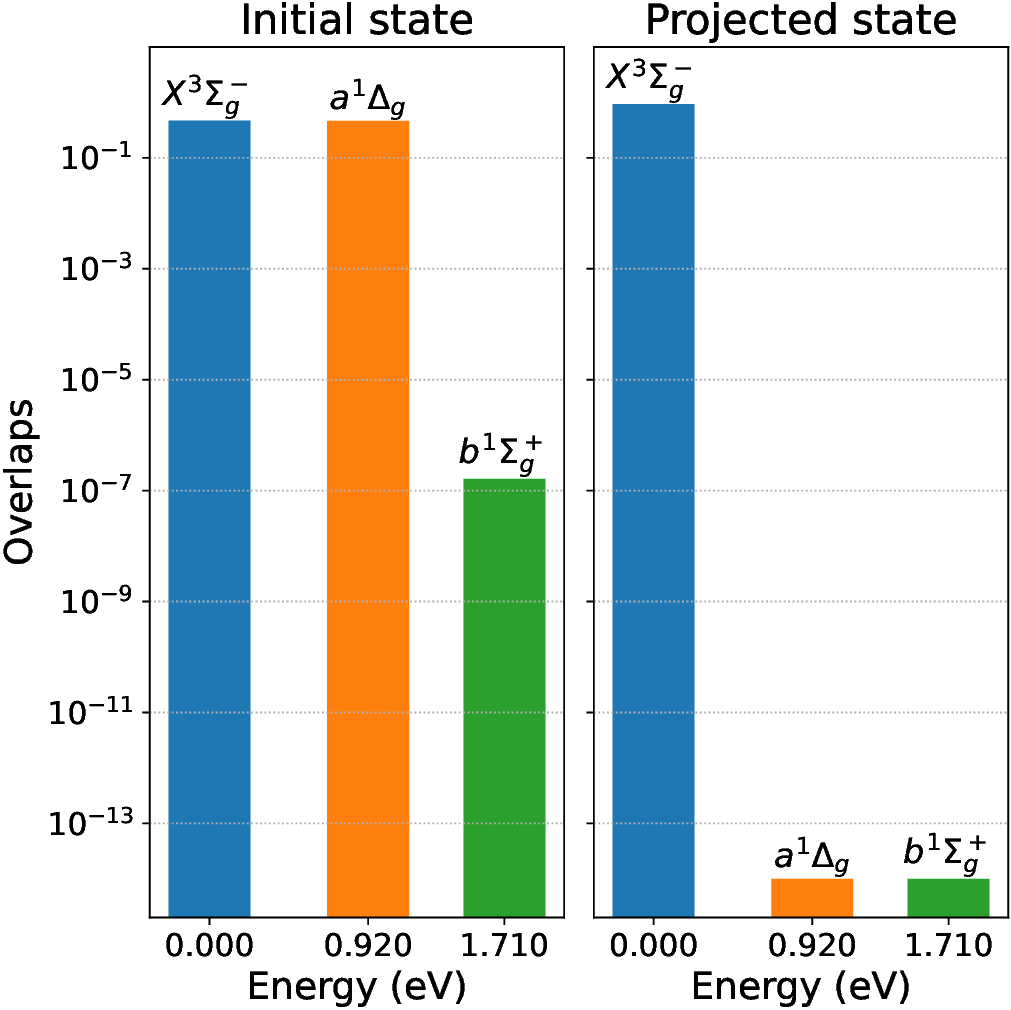}
    \caption{Overlap of the initial guess with the singlet ($a^1\Delta_g$, $b^1\Sigma_g^+$) and triplet ($X^3\Sigma_g^-$) states of O$_2$ before and after applying the $\hat{P}_{S,M_S}$ spin-symmetry projection.}
    \label{fig:O2_states}
\end{figure}

Within this active space, the initial state prepared using a hardware-efficient variational ansatz~\cite{Kandala2017} does not respect spin symmetry and produces a substantial overlap with both the triplet ground state and at least one singlet excited state.
The numerical results are presented in Fig.~\ref{fig:O2_states}.
In particular, we observed comparable amplitudes in the ${}^3\Sigma_g^-$ and $a^1\Delta_g$ states, resulting in a QPE success probability of only $p_\text{QPE} \approx 0.47$.

Applying a total spin projector $\hat{P}_{S,M_S}$ with $S=1$ and $M_S=0$ substantially suppresses singlet contamination and increases the overlap with the triplet manifold.
After symmetry filtering, the QPE success probability increases to $p^\text{filtered}_\text{QPE} \approx 0.94$

We also estimated the QPE cost for the active space Hamiltonian.
The qubitization-based QPE requires approximately $8$ ancilla qubits to handle the energy with chemical precision ($\sim 1$ mHa).
Then, the total non-Clifford cost is on the order of $10^{7}$ T gates.
By contrast, the symmetry projector requires only $\sim 1.6 \times 10^4$ T gates, which is almost $3$ orders of magnitude less.
The intrinsic probability of success of $\hat{P}_{S,M_S}$ is $\approx 0.30$.
If one wishes to deterministically prepare the filtered state, one round of AA suffices to increase the success probability to $\approx 0.97$, thereby increasing the total T cost of the projection to $\sim 4.8 \times 10^4$, which remains negligible relative to the QPE cost.

\textbf{Trimethylenemethane (TMM).}
TMM exhibits a well-studied singlet–triplet gap in the order of tens of kcal/mol. 
The ground state is a triplet ${}^3A'_2/{}^3B_2$, and the first excited state is a degenerate pair of singlets $1{}^1B_2$ and $1{}^1A_1$, located approximately $1.17$ eV above the ground state~\cite{Slipchenko2003}. 
We constructed a CAS($4$,$4$) Hamiltonian in a cc-pVTZ basis, where the active orbitals spanned the full carbon $\pi$-system.
The molecular orbitals were optimized using a triplet CASSCF calculation.

\begin{figure}
    \centering
    \includegraphics[width=1\linewidth]{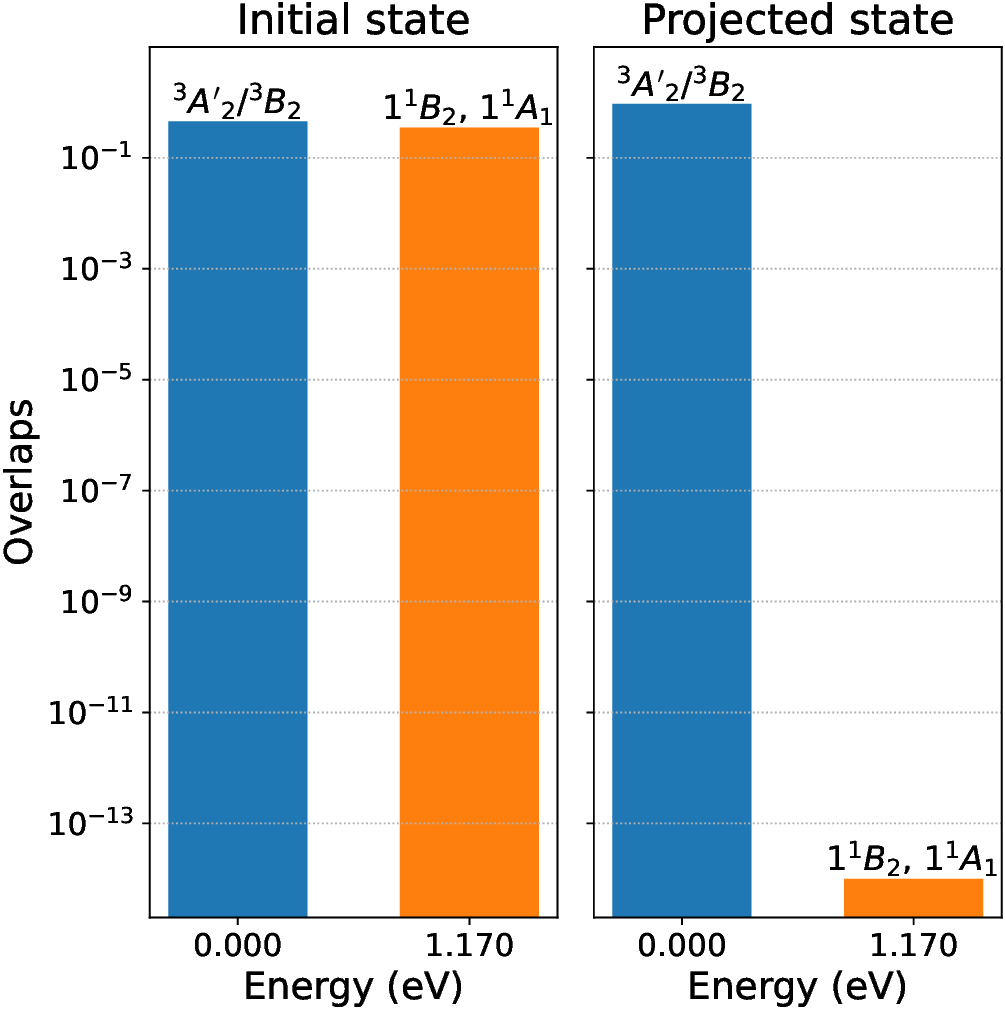}
    \caption{Overlap of the initial guess with the singlet ($1{}^1B_2$, $1{}^1A_1$) and triplet (${}^3A'_2/{}^3B_2$) states of TMM before and after applying the $\hat{P}_{S,M_S}$ spin-symmetry projection.}
    \label{fig:TMM_states}
\end{figure}

Using a hardware-efficient VQE ansatz to prepare an initial state nominally targeting the triplet ground state ${}^3A'_2/{}^3B_2$ results in noticeable contamination from the singlet states $1{}^1B_2$ and $1{}^1A_1$ (see Fig.~\ref{fig:TMM_states}).
The resulting QPE success probability was $p_\text{QPE} \approx 0.46$.
Applying the spin projector $\hat{P}_{S,M_S}$ with $S=1$ and $M_S=0$ suppresses unwanted singlet components, increases the overlap with the triplet sector, and improves the QPE success probability to $p^\text{filtered}_\text{QPE} \approx 0.94$.

We also estimated the QPE cost for the active space Hamiltonian.
For the CAS($4$,$4$) Hamiltonian, qubitization-based QPE with chemically accurate energy resolution requires approximately $8$ ancilla qubits on the order of $10^{7}$ T gates.
In contrast, the spin projector requires approximately $7.7\times10^3$ T gates, which is more than $3$ orders of magnitude lower than QPE.
The intrinsic success probability of the projection is $\approx 0.29$; therefore, a single step of the AA is sufficient to increase the probability above $0.98$.
This increases the total T cost of the projector to $\sim 2.3\times10^4$, which is negligible compared with the cost of the subsequent QPE.

\begin{table*}[tbp!]
\centering
\caption{Estimated overlaps using Eq.~\eqref{eq:overlap_estimation} and total T gate costs for FeMoco active space models using amplitude amplification (AA) with the total-spin projector $P_{S,M_S}$. 
The estimated T gate cost of QPE for FeMoco is $\sim 10^{10}$, shown here as a reference baseline.
The overlap $p$ is the probability that the input broken-symmetry Slater determinant lies in the target $(S,M_S)$ subspace prior to amplification.}
\label{tab:femoco}
\begin{tabular}{lcccccc}
\hline\hline
\textbf{Model} & \textbf{Electrons} $N_\text{elec}$ & \textbf{Target Spin} $S$ & \textbf{$M_S$} & 
\textbf{Overlap $p$} & \textbf{Queries (AA)} & \textbf{T Gates (Total)} \\
\hline
FeMoco ($54$ orbitals, Reiher) & $54$ & $0$ & $0$ & $0.036$ & 7 & $1.5\times 10^{7}$ \\
FeMoco ($76$ orbitals, Li), $S=3/2$ & $113$ & $3/2$ & $1/2$ & $0.068$ & 5 & $1.1\times 10^{7}$ \\
FeMoco ($76$ orbitals, Li), $S=1/2$ & $113$ & $1/2$ & $1/2$ & $0.035$ & 7 & $1.5\times 10^{7}$ \\
\hline\hline
\end{tabular}
\end{table*}

\textbf{Implications for FeMoco.}
The FeMoco cluster presents an extreme case with strong static correlation. 
Even in reduced active space models ($54$~\cite{Reiher2017} or $76$~\cite{Li2019} spatial orbitals), multiple states can lie within chemically relevant energy differences. 
In such cases, increasing the initial overlap with the correct irreducible subspace prior to the phase estimation can significantly reduce the number of QPE repetitions required. 

Recent large-scale classical calculations on the $76$-orbital FeMoco model indicate that a single broken-symmetry Slater determinant can already carry a substantial amplitude in an accurate ground state approximation.
In particular, overlaps between the dominant broken-symmetry determinant $|\psi_{\mathrm{BS}}\rangle$ and a highly converged UDMRG MPS (used as an approximate ground state) reach $|\langle \psi_{\mathrm{BS}}|\phi_0\rangle| \approx 0.45$--$0.47$ for representative low-energy spin isomers, while the next-largest determinant coefficient is below $0.1$~\cite{Zhai2026}.
If one uses $|\psi\rangle = |\psi_{\mathrm{BS}}\rangle$ as an input guess for QPE, this corresponds to an estimated success probability $p_{\mathrm{QPE}} \approx |\langle \psi_{\mathrm{BS}}|\phi_0\rangle|^2 \approx 0.2$.

However, broken-symmetry determinants are generally not eigenstates of $\hat S^2$ and therefore contain components in multiple total-spin sectors. 
Consequently, a spin projector can serve as a natural filter that removes contributions outside the target $(S,M_S)$ sector and concentrates the remaining weight into the desired irreducible subspace.

The QPE cost for the FeMoco system can be estimated as $\sim 10^{10}$ T gates by using tensor hypercontraction~\cite{Lee2021}.
At the same time, $P_{S,M_S}$ projector realized using GQSP would require $\sim 10^6$ T gates, which is $4$ orders of magnitude less than QPE cost.

In reality, several orders will be taken by the AA, which can be roughly estimated if we assume a uniform superposition of the states of $N_\text{elec}$ electrons, total spin $S$ and fixed $M_S$.
The number of multiplets is 
\begin{equation}
    n(S) = \binom{N_\text{elec}}{\frac{N_\text{elec}}{2}+S} - \binom{N_\text{elec}}{\frac{N_\text{elec}}{2}+S+1}.
\end{equation}
Thus, the expected squared overlap of a random state in the $M_S=S$ sector is
\begin{equation}
\label{eq:overlap_estimation}
    p_\text{min} = \frac{n(S)}{\binom{N_\text{elec}}{\frac{N_\text{elec}}{2}}}.
\end{equation}
Therefore, the estimated probability is approximately $0.03$--$0.07$ depending on the FeMoco model (see Table~\ref{tab:femoco}).
This probability resulted in $5$-$7$ steps of AA.
Then, the GQSP implementation of $P_{S,M_S}$ projector together with the AA costs approximately $\sim 10^7$ of the T gates, which is still a $3$ order advantage over QPE itself.

Such symmetry filtering has been shown to significantly increase overlap in smaller systems, such as O$_2$ and TMM.
By analogy, one may therefore expect that the same strategy can substantially enhance the initial overlap for FeMoco, thereby reducing the number of required QPE repetitions.
Taken together, these considerations suggest that the state preparation problem for FeMoco, while still nontrivial, may no longer be the dominant bottleneck when physically motivated broken-symmetry references are combined with symmetry projection and AA.

\subsection{Limitations and outlook}
\label{subsec:limitations}

Although the projector-based filtering approach improves the effective overlap with the target eigenstate prior to QPE, several limitations remain. 
First, the probability of the success of the projection step may be low when the initial state carries only a modest component in the desired symmetry sector. 
Although AA can boost this probability, the number of amplification rounds scales as inverse overlap. 
In cases in which the trial state is poorly chosen, the overhead associated with amplitude amplification may become significant. 
This highlights the importance of constructing a good initial ansatz, either from classical multireference methods or using adaptive quantum state preparation strategies.

Although we focused on continuous symmetries ($\hat{N}$, $\hat{S}_z$, $\hat{S}^2$), the same GQSP/GQSVT frameworks can be applied to finite molecular point groups when the symmetry actions $U(g)$ are available. 
For any irreducible representation $\Gamma$ of finite group $G$, the projector
\begin{equation}
P_\Gamma = \frac{d_\Gamma}{|G|}\sum_{g\in G} \chi_\Gamma(g)^{*}\,U(g)
\label{eq:PGamma}
\end{equation}
selected the $\Gamma$-subspace. 
For abelian groups, all irreducible representations are one-dimensional and \eqref{eq:PGamma} reduces to a weighted average of symmetry operations, which is directly compatible with LCU, GQSP, and GQSVT implementations. 
If $G$ decomposes as a direct or semidirect product (such as $\mathrm{D}_{3h} \cong \mathrm{C}_{3v} \times \mathrm{C}_{s}$ and $\mathrm{C}_{3v} \cong \mathrm{C}_{3} \rtimes \mathrm{C}_{s}$), the corresponding projectors factorize as $P_{\mathrm{D}_{3h}} = P_{\mathrm{C}_{3v}} P_{\mathrm{C}_{s}}$, allowing symmetry filtering to be applied sequentially. 
Another method involves constructing each symmetry operator and projecting its specific eigenvalue using the Lagrange interpolation and GQSVT.

The projector and AA can be merged into a single deterministic procedure using a double-bracket symmetry refinement. 
Given Hermitian generator $\hat{O}$, the evolution
\begin{equation}
\ket{\psi(\tau)} \propto \exp\!\bigl(-\tau [\hat{O},[\hat{O},\cdot]]\bigr)\ket{\psi(0)}
\end{equation}
suppresses the components outside the target symmetry sector, while leaving the desired subspace invariant. 
A discretized implementation in terms of the controlled applications of $\hat{O}$ yields deterministic symmetry purification at the cost of an additional depth.

%% file: Sections/05_Conclusions.tex
\section{Conclusion} 
\label{sec:conclusion}

In this study, we constructed symmetry projectors for quantum state preparation based on LCU, GQSP, and GQSVT, together with circuit realizations and parameter choices that achieve a prescribed projection accuracy. 
To characterize their performance, we analyzed the numerical accuracy, gate-precision sensitivity, and asymptotic resource scaling, and translated them into concrete CNOT and T gate estimates. 
The comparison isolated the main trade-offs among the three approaches.
LCU was the most rotation-robust and simplest to parameterize.
GQSP achieved similar leading-order T gate scaling with minimal ancilla overhead (one qubit) but required slightly tighter rotations.
GQSVT attained exact polynomial filtering on discrete spectra but was more phase-sensitive and carried larger constant factors in both entangling and non-Clifford resources.

For $\hat{N}$ and $\hat{S}_z$, both LCU and GQSP exhibited quasi-quadratic scaling in the system size, with T gate counts that grew as $N_{\mathrm{SO}}^2 \log N_{\mathrm{SO}}$.
The two methods differed only in small additive terms, with LCU maintaining a slight advantage in the total T cost. 
GQSVT showed a steeper entangling cost (CNOT scaling roughly as $N_{\mathrm{SO}}^2 \log N_{\mathrm{SO}}$) and a larger T gate prefactor, making it several times more expensive at a fixed accuracy, although it can be competitive at a very small $N_{\mathrm{SO}}$.
For total-spin filtering, the composition $\hat{P}_{S,M_S}=\hat{P}_{M_S}\hat{P}_S\hat{P}_{M_S}$ introduces the cost of preparing fixed states $S$ into the same quasi-quadratic T gate scaling class as $\hat{N}$ and $\hat{S}_z$. 
Importantly, the $\beta$ integration for $\hat{P}_S$ depends weakly on the target spin $S$ and is independent of $N_{\mathrm{SO}}$.

Our precision analysis provided practical synthesis targets.
The LCU remained accurate with a single-qubit rotation precision of approximately $10^{-1}$. 
For GQSP, signal-operator angles of approximately $10^{-0.8}$ and QSP phase rotations better than $10^{-1}$ were typically sufficient, with $10^{-2}$ being a safe choice. 
GQSVT required substantially tighter phase control, with a rotation precision near $10^{-3.7}$ whereas PREP-type data paths tolerate $\sim10^{-2}$. 
These thresholds were mapped directly to the T gate counts via standard Clifford+T synthesis, implying only a logarithmic overhead per additional decade of precision.

We demonstrated that symmetry projectors act as effective QPE filters in representative molecular systems. 
In O$_2$ and TMM, symmetry filtering increased the QPE success probability from $\sim0.46$ to $\sim0.94$. The projector costs remained roughly $3$ orders of magnitude below the subsequent QPE step.  
Even for FeMoco-scale active spaces, the total cost of projection plus AA remained $\sim10^7$ T gates versus $\sim10^{10}$ for a single qubitization-based QPE with chemical accuracy, yielding a substantial net reduction in the number of repetitions required to resolve the desired eigenvalue.

Overall, symmetry filtering is a low-cost, accuracy-preserving preprocessing step that scales favorably with the system size and materially improves the QPE success probability. 
Given its modest ancilla footprint, straightforward parameterization, and clear synthesis targets, we expect projector-based state preparation to become a standard component of fault-tolerant quantum simulation workflows, especially for strongly correlated molecules where resolving the correct symmetry sector is essential.

%% file: Sections/06_Appendix.tex
\appendix

\section{Computational primitives}
\label{appendix:primitives}

In this appendix, we summarize the computational primitives used throughout this study: LCU, GQSP, and its singular-value variant GQSVT.

\subsection{Linear combination of unitaries (LCU)}
\begin{figure}[!h]
    \centering
    \begin{quantikz}[column sep=0.5cm, row sep={0.5cm}]
        \lstick{$|0\rangle$} & \qwbundle{a} & \gate{\operatorname{PREP}} & \gate[2]{\operatorname{SELECT}} & \gate{\operatorname{PREP}^\dagger} & \meter{|0\rangle}\\
        \lstick{$|\psi\rangle$} & \qw & \qw & \qw & \arrow[r] & \rstick{$A/\alpha|\psi\rangle$}
    \end{quantikz}
    \caption{LCU circuit preparing an $(\alpha,a,0)$ block encoding of $A$. The state $A/\alpha|\psi\rangle$ is up to corresponding normalization.}
    \label{fig:LCU}
\end{figure}
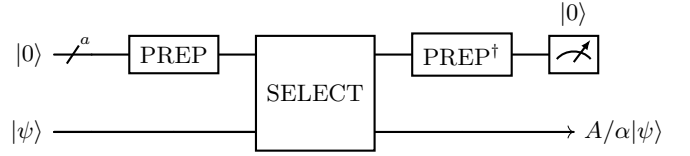

Let operator $A$ be decomposed into a linear combination of unitaries:
\begin{equation}
\label{eq:LCU}
    A = \sum_{k=0}^{L-1} w_k\, U_k,\qquad \alpha = \sum_k |w_k|.
\end{equation}
Map $|\psi\rangle \mapsto (A/\alpha)|\psi\rangle$ can be realized using an ancilla register by defining
\begin{align}
    \mathrm{PREP}:&\quad |0\rangle \mapsto |\omega\rangle = \sum_{k=0}^{L-1}\sqrt{|w_k|/\alpha}|k\rangle\,\\
    \mathrm{SELECT}:&\quad |k\rangle |\psi\rangle \mapsto |k\rangle U_k\big|\psi\rangle,
\end{align}
where $\mathrm{PREP}$ encodes the coefficients $w_k$ in Eq.~\eqref{eq:LCU}, while $\mathrm{SELECT}$ applies an indexed unitary to the working register.
The composite circuit 
\begin{equation}
    V_A \;=\; (\mathrm{PREP}^\dagger\otimes I)\,\mathrm{SELECT}\,(\mathrm{PREP}\otimes I)
\end{equation}
satisfies
\begin{equation}
    \big(\langle 0|\otimes I\big)V_A\big(|0\rangle\otimes I\big) = A/\alpha.
\end{equation}
Thus, $V_A$ is an $(\alpha,a,0)$ block encoding of operator $A$ because it prepares this operator scaled by $\alpha$ in the $\{|0\rangle, |0\rangle\}$ block of matrix $V_A$ with $0$ error using $a=\lceil\log_2 L\rceil$ ancilla qubits.
The LCU circuit is illustrated in Fig.~\ref{fig:LCU}.

The postselection success probability for the input $|\psi\rangle$ is
\begin{equation}
    p_{\mathrm{LCU}}(|\psi\rangle) = \big\| (A/\alpha)|\psi\rangle \big\|_2^2 \le \|A\|_\text{LCU}^2/\alpha^2 \le 1,
\end{equation}
where $\|A\|_{\mathrm{LCU}} = \min_{\{ w_k\}} \sum_k |w_k|,$ is the LCU norm, which corresponds to the smallest possible sum of the LCU coefficients through all possible decompositions. 
The probability of success can be increased to nearly unity through AA.

The PREP can be synthesized via a binary-tree ladder of controlled rotations~\cite{binary_tree} by applying $O(L)$ controlled rotations to $\lceil\log_2 L\rceil$ ancilla qubits. 
To achieve precision in the overall state preparation, $\epsilon$, each rotation must be accurate to $O(\epsilon/L)$, which produces $O(\log(L/\epsilon))$ T gates per rotation.

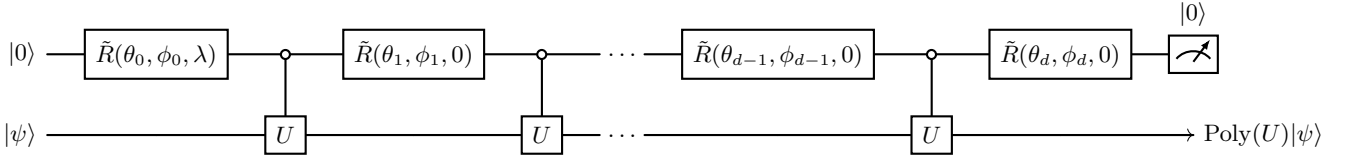
\begin{figure*}[tbp]
  \centering
  \begin{quantikz}[column sep=0.5cm, row sep={0.5cm}]
    \lstick{$\ket{0}$}
    & \gate{\tilde{R}(\theta_0, \phi_0, \lambda)}
    & \octrl{1}
    & \gate{\tilde{R}(\theta_1, \phi_1, 0)}
    & \octrl{1}
    & \ \ldots \
    & \gate{\tilde{R}(\theta_{d-1}, \phi_{d-1}, 0)}
    & \octrl{1}
    & \gate{\tilde{R}(\theta_{d}, \phi_{d}, 0)}
    & \meter{|0\rangle}\\
    \lstick{$\ket{\psi}$}
    &
    & \gate{U}
    &
    & \gate{U}
    & \ \ldots \
    &
    & \gate{U}
    & \arrow[r] & \rstick{$\operatorname{Poly}(U)|\psi\rangle$} \\
  \end{quantikz}
    \caption{GQSP circuit. The circuit comprises one signal processing qubit and a system register on which $\operatorname{Poly}(U)$ acts. The state $\operatorname{Poly}(U)|\psi\rangle$ is up to corresponding normalization.}
  \label{fig:GQSP_cirsuit}
\end{figure*}

However, the T gate count of the PREP can be improved using a quantum read-only memory (QROM) scheme~\cite{QROM}. 
The idea behind QROM is that the square roots of the normalized LCU coefficients, $\sqrt{|w_i|/\alpha}$ are stored as classical binary data in the table of $L$ entries.
These values are loaded into the quantum state as follows:
\begin{equation}
    |0\rangle|0\rangle \xrightarrow{\mathrm{QROM}}
    \sum_k C_k|k\rangle\,|d_k\rangle,
\end{equation}
where $d_k$ denotes the a classical binary data associated with index $k$.
This allows the preparation of a state
\begin{equation}
    |0\rangle^{\otimes(1 + 2\log L + 2\mu)} \to
    \sum_{k=0}^{L-1} \sqrt{\frac{\tilde{w}_k}{\alpha}}|k\rangle|\text{temp}_k\rangle,
\end{equation}
with a $\mu$-bit approximation $\tilde{w}_k/\alpha$ and garbage state $|\mathrm{temp}_k\rangle$. 
Although this method uses $1+2\lceil\log L\rceil+2\mu$ ancilla qubits, it requires only $4L + O(\log(1/\epsilon))$ T gates with $\mu=O(\log(1/\epsilon))$, which provide linear scaling in $L$ with a precision-independent prefactor.

A basic SELECT applies indexed multi-controlled $U_k$ gate without optimization costs $O(L\log L)$ T gates. 
Using clean ancillas and gate merging~\cite{QROM} reduces the multi-control overhead to $4L-4$ T-gates. 
These counts capture only multi-control scaffolding.
If $U_k$ are rotation unitaries, their own synthesis costs add and can dominate, whereas Pauli strings may be generated at a lower cost.

\subsection{Generalized quantum signal processing (GQSP)}

GQSP enables the implementation of an arbitrary complex polynomial $\operatorname{Poly}(z)$ acting on a unitary $U$~\cite{GQSP}. 
Unlike the conventional QSP, which is restricted to real-valued and fixed parity polynomials, GQSP allows for arbitrary complex-valued and mixed parity polynomial transformations.
The only limitation is that this polynomial must obey $|\operatorname{Poly}(z)| \leq 1$ in the complex unit circle $|z|=1$.

The controlled signal operator is defined as
\begin{equation}
\label{eq:GQSP_controlledU}
\text{C}U = \ket{0}\!\bra{0} \otimes U + \ket{1}\!\bra{1} \otimes I,
\end{equation}
which applies $U$ conditioned on the ancilla as $\ket{0}$. 
The GQSP sequence alternates $\mathrm{C}U$ with single-qubit rotations $\tilde{R}(\theta,\phi,\lambda)$ in the ancilla:
\begin{equation}
\label{eq:GQSP_sequence}
\begin{bmatrix}
\operatorname{Poly}(U) & \cdot \\
\cdot & \cdot
\end{bmatrix}
\;=\;
\prod_{k=1}^{d}
\Big(\tilde{R}(\theta_k, \phi_k,0) \,
    \text{C}U
\Big)\tilde{R}(\theta_0, \phi_0,\lambda),
\end{equation}
with phases $\{\theta_k,\phi_k,\lambda\}$ selected to realize a degree $d$ polynomial~\cite{GQSP, yamamoto2024robustanglefindinggeneralized, Berntson2025}. 
This yields the block encoding of $\operatorname{Poly}(U)$ (see Fig.~\ref{fig:GQSP_cirsuit}). 
The rotation is defined as
\begin{equation}
    \tilde{R}(\theta, \phi,\lambda) 
    = 
    \begin{bmatrix}
    e^{i(\lambda+\phi)} \cos(\theta) & e^{i\phi} \sin(\theta) \\
    e^{i\lambda} \sin(\theta) & - \cos(\theta)
    \end{bmatrix}.
\end{equation}

In general, the GQSP protocol is probabilistic because the $\operatorname{Poly}(U)$ transformation may not be unitary.
The measurement of the ancilla in $\ket{0}$ is successful with probability
\begin{equation}
p_{\mathrm{GQSP}}(\ket{\psi}) \;=\;
\big\| \operatorname{Poly}(U)\ket{\psi} \big\|_2^2.
\end{equation}
Employing AA, the procedure can be made nearly deterministic, thereby ensuring the high-fidelity realization of the target polynomial transformation.

\begin{figure*}[tbp]
  \centering
  \begin{quantikz}[column sep=0.5cm, row sep={0.5cm}]
    \lstick{$\ket{0}$}
    & \gate{\tilde{R}(\theta_0, \phi_0, \lambda)}
    &\gategroup[3,steps=4,style={dashed,rounded
corners,fill=blue!20, inner
xsep=2pt},background,label style={label
position=below,anchor=north,yshift=-0.2cm}]{{\text{Controlled-}$Q_A$}} 
    & \octrl{1}
    &
    & \gate{Z}
    & \ \ldots \
    & \gate{\tilde{R}(\theta_{d}, \phi_{d}, 0)}
    & \meter{|0\rangle}\\
    \lstick{$\ket{0}$}
    & \qwbundle{a}
    & \gate{\operatorname{PREP}}
    & \gate[2]{\operatorname{SELECT}}
    & \gate{\operatorname{PREP}^\dagger}
    & \octrl{-1}
    & \ \ldots \
    &
    & \meter{|0\rangle}\\
    \lstick{$\ket{\psi}$}
    & \phantomgate{really wide gate}
    & \qw
    & \qw
    & \qw
    & \qw
    & \ \ldots \
    & \arrow[r] & \rstick{$\operatorname{Poly}(A/\alpha)|\psi\rangle$}
  \end{quantikz}
    \caption{GQSVT circuit. The circuit comprises three registers: the first one qubit register performs signal processing; the second $a$ qubits register performs the block encoding; and the third qubits register holds the state on which the operator $\operatorname{Poly}(A/\alpha)$ acts. The state $\operatorname{Poly}(A/\alpha)|\psi\rangle$ is up to corresponding normalization. The blue frame highlights the controlled-$Q_A$ qubitization operator.}
  \label{fig:GQSVT_cirsuit}
\end{figure*}
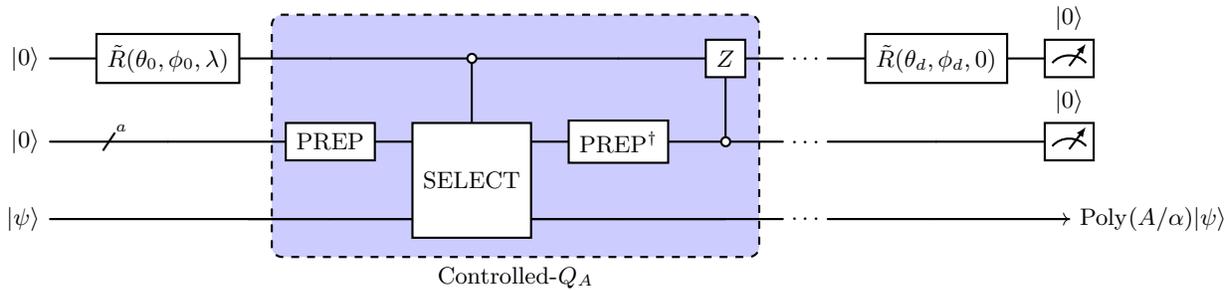

\subsection{Generalized quantum singular-value transformation (GQSVT)}

The LCU construction introduced above naturally leads to a qubitization technique~\cite{qubitization}.
Given a block encoding $V_A$ of $A/\alpha$, qubitization embeds this operator into a larger unitary, whose action is reduced to two–dimensional invariant subspaces.
Concretely, define
\begin{equation}
\label{eq:qubitization}
Q_A \;=\;
(2 \ket{0}\!\bra{0} \otimes I - I)\, V_A.
\end{equation}
Within each invariant subspace, $Q_A$ acts as an $SU(2)$ rotation.
If $\sigma_i$ are singular values of $A$, then the eigenphases $\pm \theta_i$ of $Q_A$ satisfy
\begin{equation}
    \cos \theta_i = \sigma_i / \alpha,
\end{equation}
such that the spectrum of $A$ is encoded at the rotation angles of $Q_A$.

Using $Q_A$ as the signal operator in the GQSP framework yields the generalized quantum singular value transformation (GQSVT)~\cite{GQSVT}, in which polynomial transformations act directly on the singular values of $A$.
The resulting sequence has the same form as Eq.~\eqref{eq:GQSP_sequence} with controlled-$Q_A$ instead of controlled-$U$.
If $A$ is Hermitian, then the procedure implements $\operatorname{Poly}(A/\alpha)$.
The GQSVT circuit is illustrated in Fig.~\ref{fig:GQSVT_cirsuit}.

Therefore, the GQSVT framework provides a unified method for constructing projectors, filters, and other spectral transformations for any block-encoded operator without assuming normality.
The probability of success for an input state $\ket{\psi}$ is
\begin{equation}
p_{\mathrm{GQSVT}}(\ket{\psi}) \;=\;
\big\| \operatorname{Poly}(A/\alpha)\ket{\psi} \big\|_2^2,
\end{equation}
and can be boosted using AA, similar to GQSP.